\documentclass[aps,prb,reprint,amsfonts,amssymb,amsmath,superscriptaddress,longbibliography,floatfix]{revtex4-2}

\pdfoutput=1

\usepackage{amssymb,amsmath,bm}
\usepackage{fullpage}
\usepackage{graphicx}
\usepackage{bibunits}
\usepackage{amsmath}
\usepackage{wrapfig}
\usepackage{floatflt}
\usepackage{color,soul}
\usepackage{xspace}
\usepackage{dsfont}
\usepackage[official,right]{eurosym}
\usepackage[top=0.8in, bottom=0.8in, left=0.75in, right=0.75in]{geometry}
\usepackage[utf8]{inputenc}
\usepackage[colorlinks=true,urlcolor=blue,citecolor=blue,linkcolor=blue,bookmarks=false]{hyperref}
\usepackage[flushleft]{threeparttable}
\usepackage{array,booktabs,makecell}
\usepackage{tabularx,array}
\usepackage{cleveref}
\usepackage{csquotes}
\MakeOuterQuote{"}
\newcolumntype{Y}{>{\centering\arraybackslash}X}

\newcommand{\pbwo}{Pr$_{3}$BWO$_9$\xspace}
\newcommand{\nbwo}{Nd$_{3}$BWO$_9$\xspace}
\newcommand{\pebwo}{Pr$_{3}\,^{11}$BWO$_9$\xspace}

\newcommand{\ket}[1]{\left| #1 \right\rangle}
\newcommand{\bra}[1]{\left\langle #1 \right|}

\newcommand{\be}{\begin{equation}}
\newcommand{\ee}{\end{equation} }
\newcommand{\bea}{\begin{eqnarray} }
\newcommand{\eea}{\end{eqnarray} }

\begin{document}

\title{Excitation Spectrum and Spin Hamiltonian of the Frustrated Quantum Ising Magnet \pbwo}
\author{J.~Nagl}
\email{jnagl@ethz.ch}
\affiliation{Laboratory for Solid State Physics, ETH Z{\"u}rich, 8093 Z{\"u}rich, Switzerland}

\author{D.~Flavi{\'a}n}
\affiliation{Laboratory for Solid State Physics, ETH Z{\"u}rich, 8093 Z{\"u}rich, Switzerland}

\author{S.~Hayashida}
\affiliation{Laboratory for Solid State Physics, ETH Z{\"u}rich, 8093 Z{\"u}rich, Switzerland}
\affiliation{Present address: Max-Planck-Institut f\"ur Festk\"orperforschung, Heisenbergstraße 1, 70569 Stuttgart, Germany}

\author{K.~Yu.~Povarov}
\affiliation{Dresden High Magnetic Field Laboratory (HLD-EMFL) and W\"urzburg-Dresden Cluster of Excellence ct.qmat, Helmholtz-Zentrum Dresden-Rossendorf (HZDR), 01328 Dresden, Germany}

\author{M.~Yan}
\affiliation{Laboratory for Solid State Physics, ETH Z{\"u}rich, 8093 Z{\"u}rich, Switzerland}

\author{N.~Murai}
\affiliation{J-PARC Center, Japan Atomic Energy Agency, Tokai, Ibaraki 319-1195, Japan}
\author{S.~Ohira-Kawamura}
\affiliation{J-PARC Center, Japan Atomic Energy Agency, Tokai, Ibaraki 319-1195, Japan}

\author{G.~Simutis}
\affiliation{Laboratory for Neutron and Muon Instrumentation, Paul Scherrer Institut, CH-5232 Villigen-PSI, Switzerland}

\author{T.J.~Hicken}
\affiliation{Laboratory for Muon Spin Spectroscopy, Paul Scherrer Institut, 5232 Villigen, Switzerland}

\author{H.~Luetkens}
\affiliation{Laboratory for Muon Spin Spectroscopy, Paul Scherrer Institut, 5232 Villigen, Switzerland}

\author{C.~Baines}
\affiliation{Laboratory for Muon Spin Spectroscopy, Paul Scherrer Institut, 5232 Villigen, Switzerland}

\author{A. Hauspurg}
\affiliation{Dresden High Magnetic Field Laboratory (HLD-EMFL) and W\"urzburg-Dresden Cluster of Excellence ct.qmat, Helmholtz-Zentrum Dresden-Rossendorf (HZDR), 01328 Dresden, Germany}
\affiliation{Institut f\"ur Festk\"orper- und Materialphysik, Technische Universit\"at Dresden, 01062 Dresden, Germany}

\author{B. V. Schwarze}
\affiliation{Dresden High Magnetic Field Laboratory (HLD-EMFL) and W\"urzburg-Dresden Cluster of Excellence ct.qmat, Helmholtz-Zentrum Dresden-Rossendorf (HZDR), 01328 Dresden, Germany}
\affiliation{Institut f\"ur Festk\"orper- und Materialphysik, Technische Universit\"at Dresden, 01062 Dresden, Germany}

\author{F. Husstedt}
\affiliation{Dresden High Magnetic Field Laboratory (HLD-EMFL) and W\"urzburg-Dresden Cluster of Excellence ct.qmat, Helmholtz-Zentrum Dresden-Rossendorf (HZDR), 01328 Dresden, Germany}
\affiliation{Institut f\"ur Festk\"orper- und Materialphysik, Technische Universit\"at Dresden, 01062 Dresden, Germany}

\author{V.~Pomjakushin}
\affiliation{Laboratory for Neutron Scattering and Imaging, Paul Scherrer Institut, 5232 Villigen,  Switzerland}

\author{T.~Fennell}
\affiliation{Laboratory for Neutron Scattering and Imaging, Paul Scherrer Institut, 5232 Villigen, Switzerland}

\author{Z.~Yan}
\affiliation{Laboratory for Solid State Physics, ETH Z{\"u}rich, 8093 Z{\"u}rich, Switzerland}

\author{S.~Gvasaliya}
\affiliation{Laboratory for Solid State Physics, ETH Z{\"u}rich, 8093 Z{\"u}rich, Switzerland}

\author{A.~Zheludev}
\email{zhelud@ethz.ch}
\homepage{http://www.neutron.ethz.ch/}
\affiliation{Laboratory for Solid State Physics, ETH Z{\"u}rich, 8093 Z{\"u}rich, Switzerland}

\date{\today}

\begin{abstract}
	We present a thorough experimental investigation on {\it single crystals} of the rare-earth based frustrated quantum antiferromagnet \pbwo, a purported spin-liquid candidate on the breathing kagome lattice. This material possesses a disordered ground state with an unusual excitation spectrum involving a coexistence of sharp spin-waves and broad continuum excitations. Nevertheless, we show through a combination of thermodynamic, magnetometric and spectroscopic probes with detailed theoretical modeling that it should be understood in a completely different framework. The crystal field splits the lowest quasi-doublet states into two singlets moderately coupled through frustrated superexchange, resulting in a simple effective Hamiltonian of an Ising model in a transverse magnetic field. While our neutron spectroscopy data do point to significant correlations within the kagome planes, the dominant interactions are out-of-plane, forming frustrated triangular spin-tubes through two competing ferro-antiferromagnetic bonds. The resulting ground state is a simple quantum paramagnet, but with significant modifications to both thermodynamic and dynamic properties due to small perturbations to the transverse field Ising model in the form of hyperfine enhanced nuclear moments and weak structural disorder.
\end{abstract}

\maketitle

\section{Introduction}

Rare-earth based quantum magnets pose a promising platform for realizing complex emergent quantum phenomena. The combination of spin-orbit entangled moments with strong geometric frustration can stabilize a rich variety of entangled ground states such as spin-liquids \cite{balentsSpinLiquidsFrustrated2010}, spin ices \cite{bramwellSpinIceState2001} or multipolar phases \cite{kuramotoMultipoleOrdersFluctuations2009}, with often highly unusual excitation spectra \cite{tennantStudiesSpinonsMajoranas2019}. So far, most efforts have gone towards the study of magnetic Kramers ions on simple frustrated lattices such as the pyrochlore oxides \cite{rauFrustratedQuantumRareEarth2019} or triangular lattice delafossites \cite{schmidtYbDelafossitesUnique2021}. Although moving away from this paradigm can rapidly increase complexity, it also opens up various new pathways for realizing novel and emergent many-body physics. 

Case in point, non-Kramers ions in a low symmetry environment typically posses a quasi-doublet ground state, the physics of which can be mapped to an Ising model in a transverse magnetic field \cite{wangCollectiveExcitationsMagnetic1968, chenIntrinsicTransverseField2019}. The transverse field Ising model (TFIM) is an archetype of quantum magnetism with only a handful of experimental realizations \cite{sachdevQuantumPhaseTransitions2011a, coldeaQuantumCriticalityIsing2010}. Especially on geometrically frustrated lattices this model still holds tremendous promise, where the interplay of quantum fluctuations and extensive degeneracy promotes a plethora of new exotic phases \cite{chenIntrinsicTransverseField2019, moessnerTwoDimensionalPeriodicFrustrated2000, moessnerIsingModelsQuantum2001a}. Intrinsic realizations of the TFIM Hamiltonian have gained increasing traction in recent years, with hallmark examples such as the triangular lattice material TmMgGaO$_4$ \cite{liPartialUpUpDownOrder2020, dunNeutronScatteringInvestigation2021} or the tripod kagome magnet Ho$_3$Mg$_2$Sb$_3$O$_{14}$ \cite{dunQuantumClassicalSpin2020}. Nevertheless, good prototype materials are still relatively scarce and their experimental exploration has only just begun.

The recently discovered family $R_3$BWO$_9$ of quantum antiferromagnets poses a promising route to realizing such physics \cite{ashtarNewFamilyDisorderFree2020}. Trivalent rare-earth ions $R$ are arranged on a breathing kagome lattice in the hexagonal basal plane and the common problem of antisite disorder is suppressed by a large difference between ionic radii. All members investigated so far exhibit a significant frustration ratio $f = |\theta_{\rm CW}|/T_{\rm N} \gtrsim 10$ \cite{flavianMagneticPhaseDiagram2023a, zengLocalEvidenceCollective2021}, although our recent work points to the potential relevance of significant inter-plane correlations between the kagome layers \cite{flavianMagneticPhaseDiagram2023a}. 

The compound \pbwo seems particularly interesting in this context. It retains the largest Weiss temperature $\theta_{\rm CW} \simeq -7$ K in the family \cite{ashtarNewFamilyDisorderFree2020} and shows no signs of long-range order down to 0.1 K \cite{zengLocalEvidenceCollective2021}, while the low-symmetry environment of the non-Kramers Pr$^{3+}$ ion should allow for TFIM physics. Still, aside from a recent NMR study \cite{zengLocalEvidenceCollective2021} reporting persistent collective fluctuations with a small field-dependent excitation gap, not much is known about the physics of this material.

In this study, we present a thorough investigation of the magnetic ground state properties and excitation spectrum on single crystals of \pbwo. We utilize bulk magnetometry and inelastic neutron scattering on powders to characterize the single-ion ground state, showing that this compound can indeed be understood in the context of a frustrated TFIM Hamiltonian. Heat capacity, ultrasound and muon spin relaxation measurements are employed to map out the phase diagram, confirming a quantum disordered ground state in the entire $H-T$ phase space. Furthermore, the excitation spectrum is investigated through time of flight neutron spectroscopy on single crystals, revealing a coexistence of sharp spin-waves and broad continuum excitations. We propose a minimal model Hamiltonian capable of semi-quantitatively capturing all features of the spectrum, explaining these observations in terms of dispersive crystal field excitons partially broadened by weak structural disorder. A dominant dispersion perpendicular to the purported kagome planes confirms that - just like \nbwo \cite{flavianMagneticPhaseDiagram2023a} - this system is not a pure kagome magnet, but realizes a complex exchange topology of twisted triangular spin-tubes, moderately coupled by the frustrated planar interactions. The source of chemical disorder is investigated with powder diffraction and magnetometry measurements, pointing to a scenario where $\lesssim 2 \%$ of Pr$^{4+}$ impurities and a concomitant excess of oxygen create a distribution of crystal field splittings, with drastic effects on the thermodynamic and magnetic properties. Taken together, \pbwo turns out to be a relatively simple quantum paramagnet with a singlet ground state. Nevertheless, the interplay between frustrated Ising-like magnetic moments, local distortions brought about by structural disorder and strongly hyperfine-coupled nuclear spins gives rise to a number of unusual modifications to this fascinating model system.

\section{Methods and Material}

\subsection{The Material}
\label{ssec:mat}

\begin{figure}[tbp]
\includegraphics[scale=1]{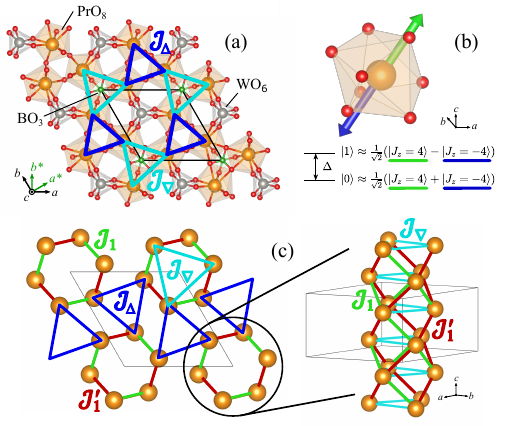}
\caption{Crystal structure and exchange pathways in \pbwo. (a) Structure in the crystallographic $ab$-plane, emphasizing the purported kagome interactions. (b) Schematic of the local magnetic environment of Pr$^{3+}$, resulting in a quasi-doublet ground state slightly split by the crystal field into superpositions of the pure Ising states $\ket{J_z = \pm 4}$. The easy axis (double arrow) is canted by $\theta \simeq 46.5^{\circ}$ from the $c$-direction. (c) The shortest bonds connect Pr ions in adjacent kagome planes through inequivalent (counter-)clockwise superexchange pathways ($\mathcal{J}_1'$)$\mathcal{J}_1$. These nearest-neighbor couplings form frustrated triangular spin-tubes along the $c$-axis.}
\label{fig:Structure}
\end{figure}

\begin{table*}[btp]
	\renewcommand{\arraystretch}{1.2}
	\caption{Potential superexchange pathways in \pbwo.}
	\begin{tabularx}{0.9\linewidth}{ccYYYYcYYY}
		\toprule\toprule
		& & \multicolumn{4}{c}{Bond Distance (\AA)} & & \multicolumn{3}{c}{Bond Angle ($^{\circ}$)} \\\cline{3-6}\cline{8-10}
		& & $r_{\mathrm{Pr-Pr}}$ & $r_{\mathrm{Pr-O_1}}$ & $r_{\mathrm{Pr-O_2}}$ & $r_{\mathrm{Pr-O_3}}$ & & $\theta_{\mathrm{Pr-O_1-Pr}}$ & $\theta_{\mathrm{Pr-O_2-Pr}}$ & $\theta_{\mathrm{Pr-O_3-Pr}}$ \\\midrule
		$\mathcal{J}_1$ & & 3.96 & 2.43 / 2.57 & 2.51 / 2.57 & - & & 104.9 & 102.5 & - \\
		$\mathcal{J}_1'$ & & 3.96 & 2.43 / 2.63 & - & 2.34 / 2.37 & & 103.0 & - & 114.5 \\
		$\mathcal{J}_\Delta$ & & 4.25 & - & 2.46 / 2.57 & - & & - & 115.1 & - \\
		$\mathcal{J}_3$ & & 4.39 & - & 2.46 / 2.51 & - & & - & 124.4 & - \\
		$\mathcal{J}_\nabla$ & & 4.94 & 2.57 / 2.63 & - & - & & 143.4 & - & - \\
		\bottomrule
	\end{tabularx}
	\label{tab:bonds}
\end{table*}

\pbwo crystalizes in a hexagonal structure with space group $P6_3$ and lattice parameters $a = 8.7210(1) \text{ \AA}$ and $c = 5.5049(1) \text{ \AA}$. Its magnetism stems from the six Pr$^{3+}$ ions in each unit cell with $4f^2$ electronic configuration. Due to the trivial site symmetry (point group $C_1$), the $J = 4$ non-Kramers multiplet is fully split by the crystal electric field (CEF) and the single-ion ground state must be a singlet. Any potential low-energy magnetic properties then result from a mixing with low-lying CEF excitations through perturbations such as superexchange. If these interactions are sufficiently strong and not exceedingly frustrated, the lowest excitation branch may condense into a soft mode, giving rise to long-range magnetic order \cite{thalmeierInducedQuantumMagnetism2023}.

As shown in Figure \ref{fig:Structure}(a), the Pr$^{3+}$ ions are arranged into a breathing-kagome lattice in the crystallographic basal plane, stacked in an AB alternating fashion. Neighboring triangles have slightly different sizes, resulting in a breathing anisotropy with inequivalent in-plane exchanges $\mathcal{J}_\Delta$ and $\mathcal{J}_\nabla$. However, the shortest Pr-O-Pr bond actually corresponds to an out of plane coupling connecting the larger $\mathcal{J}_\nabla$ triangles and has to be considered on equal footing with the kagome interactions \cite{flavianMagneticPhaseDiagram2023a} (see Table \ref{tab:bonds} for details). Due to several inequivalent superexchange pathways, this nearest-neighbor interaction must be divided into two seperate couplings $\mathcal{J}_1$ and $\mathcal{J}_1'$. These can partially release the frustration, making up twisted triangular spin-tubes along the crystallographic $c$-axis as schematically depicted in Figure \ref{fig:Structure}(c), with ($\mathcal{J}_1'$) $\mathcal{J}_1$ chains running (counter-)clockwise. While the kagome interactions must be antiferromagnetic following the Kanamori-Goodenough rules \cite{goodenoughMagnetismChemicalBond}, the smaller bond angles $\theta \sim 100^{\circ} - 115^{\circ}$ for nearest neighbors could result in either ferromagnetic (FM) or antiferromagnetic (AFM) correlations.

Previous studies reported a moderate Weiss temperature $\theta_{\rm CW} \approx -7$ K \cite{ashtarNewFamilyDisorderFree2020}, reflective of dominant antiferromagnetic interactions. In zero field, no magnetic order has been found down to dilution temperatures $T \approx 0.1$ K \cite{zengLocalEvidenceCollective2021}, resulting in a considerable frustration ratio $f = |\theta_{\rm CW}|/T_{\rm N} \gtrsim 70$. This claim must be approached with some caution though, as even a completely unfrustrated model (e.g. on a simple cubic lattice) will remain disordered if the exchanges cannot close the gap to the first CEF excitation. A recent NMR investigation \cite{zengLocalEvidenceCollective2021} has revealed persistent collective fluctuations with a small field-dependent gap of order $\sim 10$ K. Nevertheless, all the work performed so far on this system has focused exclusively on the frustrated kagome interactions. A more thorough study, including a careful treatment of the crystal electric field and the nearest-neighbor interplane interactions is still missing from the literature.

\subsection{Experimental Details}

\begin{figure}[tbp]
	\includegraphics[scale=1]{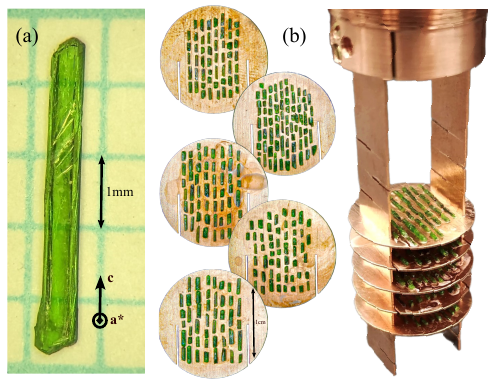}
	\caption{(a) Typical single crystal sample of \pbwo. (b) The probe used for time-of-flight neutron spectroscopy, containing a total of 282 single crystals coaligned in the $bc$ scattering plane.}
	\label{fig:Sample}
\end{figure}

Powder samples of \pbwo were prepared by solid-state reaction. High quality single-crystals were grown from these powders using a flux method analogous to \cite{majchrowskiGrowthSpectroscopicCharacterization2003}, resulting in needle-shaped green transparent crystals with six hexagonal faces and a typical mass of order $5 - 20$ mg, see Figure \ref{fig:Sample}(a). $^{11}$B-enriched samples for neutron studies, as well as non-magnetic La$_3$BWO$_9$ powders were prepared by the same methods. The structure and sample quality of all powders/single crystals was confirmed through x-ray diffraction, using Rigaku MiniFlex and Bruker APEX-II instruments respectively. Furthermore, a low temperature structural refinement was performed at the thermal neutron diffractometer HRPT \cite{FisherPhysBCM2000} at Paul Scherrer Institute (PSI, Switzerland). 10 g of \pebwo powders were sealed in a vanadium can and diffractograms were collected at $T = 1.5$ K for two different wavelengths $\lambda = 1.15 \text{ \AA}$ and $\lambda = 1.5 \text{ \AA}$.

Magnetic susceptibility $\chi = M/H$ measurements on a 3.6 mg single crystal were carried out using an Magnetic Property Measurement System (MPMS) SQUID magnetometer in a probing field $\mu_0 H = 0.1$ T along the crystallographic axes \textit{a$^*$}, \textit{b} and \textit{c}. The low-temperature magnetization $M$ was measured up to 7 T in the same experimental configuration. High-field magnetization curves were collected with a compensated pickup coil setup in pulsed fields up to 50 T at the High Magnetic Field Laboratory in Dresden \cite{SkourskiPRB832011}; the signal was calibrated to absolute units using the low-field SQUID data.
Heat capacity of a 0.16 mg single crystal sample was measured by means of relaxation calorimetry on a 14 T Physical Property Measurement System (PPMS) with a $^3$He-$^4$He dilution refrigerator insert.
Sound velocity measurements of the longitudinal $c_{33}$ mode (polarization $\mathbf{u}$ and propagation $\mathbf{k}$ along the crystallographic \textit{c}-axis) were carried out at the Dresden High Magnetic Field Laboratory using a pulse-echo method with phase-sensitive detection \cite{luthiPhysicalAcousticsSolid1967, WolfPhysBCM2001}. 36$^{\circ}$ Y-cut LiNbO$_3$ transducers were bonded to two parallel faces of a $L = 2.15$ mm long single cystal sample. Measurements were taken in the fundamental $f = 25.8$ MHz mode using a $^3$He-$^4$He dilution system in static fields up to 18 T, as well as a $^3$He cryostat in pulsed fields up to 60 T.

Muon spin relaxation ($\mu$SR) measurements on powder samples were carried out on the newly commissioned FLAME spectrometer at PSI. A pressed powder pellet was mounted on a Cu sample holder and installed in a $^3$He-$^4$He dilution refrigerator. Positively charged muons were implanted in the sample and asymmetry profiles of the subsequent muon decay were registered in the forward- and backward positron detectors. Measurements were taken in the temperature range 30 mK $\lesssim T \lesssim$ 200 K and in fields up to 0.5 T. The $\mu$SR spectra were analyzed using the
MUSRFIT \cite{suterMusrfitFreePlatformIndependent2012} software package.

Inelastic neutron scattering (INS) experiments on powders were performed at the thermal neutron triple-axis-spectrometer EIGER \cite{StuhrPhysBCM2017} at PSI. 12 g of \pebwo were sealed in an aluminum can and installed in a $^{4}$He orange cryostat. We chose a final wavelength of $k_f$= 2.66 \AA$^{-1}$ ($\lambda$ = 2.36 \AA), using a pyrolytic graphite filter to minimize higher-harmonic scattering. Energy scans at constant scattering angle $2\theta=10^\circ$ were taken at various temperatures.
Neutron spectroscopy measurements on single crystals were carried out at the cold disc-chopper spectrometer AMATERAS \cite{NakajimaJPSJS802011} in the Japan Proton Accelerator Research Complex (J-PARC). 282 hexagonal crystals were coaligned in the $(h,h,l)$ scattering plane for a total sample mass of 410 mg, depicted in Figure \ref{fig:Sample}(b). The probe was installed in a $^3$He cryostat and measurements were carried out at base temperature $T = 0.3$ K with incident energies $E_i = 5.92$ meV and 2.63 meV, providing coverage of the entire first Brillouin zone at an elastic FWHM energy resolution of $\Delta E = 146$ $\mu$eV and 45 $\mu$eV respectively. The resulting spectra were analyzed using the Horace \cite{ewingsHORACESoftwareAnalysis2016} and Sunny \cite{sunny_git} software packages.

\section{Experimental results}

\subsection{Magnetometry}

\begin{table}
	\renewcommand{\arraystretch}{1.2}
	\caption{Fit results for a Curie-Weiss analysis in the temperature range 10 K $\leq T \leq$ 40 K.}
	\centering
	\begin{tabularx}{0.45\textwidth}{Y Y Y}
		\toprule\toprule
		&  $\theta_{\rm CW}$ (K) & $ \mu_{\rm eff} $ ($\mu_{\rm B}$) \\
		\midrule
		$\textbf{H}\parallel \textit{a}^*$ & -7.74 & 2.76 \\ 
		$\textbf{H}\parallel \textit{b}$& -7.98 &  2.80 \\ 
		$\textbf{H}\parallel \textit{c}$ & -5.72 & 3.71 \\ 
		\bottomrule
	\end{tabularx}
	\label{table:Curie}
\end{table}

\begin{figure}[tbp]
\includegraphics[scale=1]{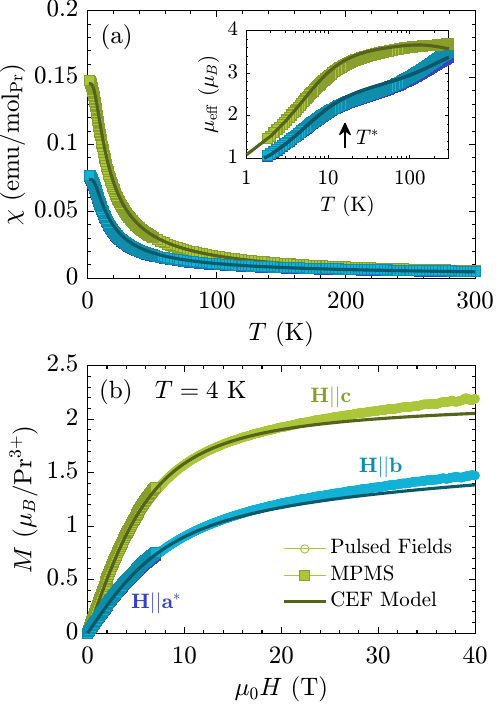}
\caption{Magnetometry measurements on single crystals. (a) Magnetic susceptibility in applied fields $\mu_0 H = 0.1$ T along three principal crystallographic axes. The in-plane response (along $\mathbf{H} \parallel a^\ast$ and $b$) is nearly axially symmetric and almost indistinguishable in the plot. The effective moment is depicted in the inset. (b) Magnetization curves for the same field orientations in both pulsed and static fields. The solid lines in all panels represent calculated values based on the crystal field model discussed in Sec. \ref{sec:GS}.}
\label{fig:Magnetometry}
\end{figure}

The bulk magnetic susceptibility $\chi$ along different crystallographic axes is shown in Figure \ref{fig:Magnetometry}(a). There are no signs of magnetic order down to 2 K and a simple fit to the inverse susceptibilities in the range $10 - 40$ K results in an average Weiss temperature $\theta_{\rm CW} \approx -7.1$ K (see Table \ref{table:Curie} for details), in agreement with previous reports \cite{ashtarNewFamilyDisorderFree2020}. As the temperature is lowered, a pronounced $\sim 2:1$ easy axis anisotropy emerges along the $c$-direction, while the planar response remains nearly axially symmetric over the entire temperature range. This should not be related directly to the single-ion anisotropy of Pr$^{3+}$ though, as the bulk susceptibility measured in experiment represents an average over the six ions in the unit cell, whose easy axes are related by $60^{\circ}$ rotations in the hexagonal plane (see Figure \ref{fig:EIGER}(c)). In the low temperature limit, the susceptibility exhibits a small downturn, which could result either from interactions or from a saturation of the large Van-Vleck susceptibility expected for a singlet ground state system.

The effective moment $\mu_{\rm eff} \propto \sqrt{\chi T}$ is depicted in the inset of Figure \ref{fig:Magnetometry}(a). It is more sensitive to small changes in the magnetism and clearly shows a hump feature at $T^{*} \sim 15$ K, below which $\mu_{\rm eff}$ begins to decline significantly. This can be understood as a depopulation of excited crystal field states and suggests the presence of at least one low-lying excitation at the scale of $T^{*}$. At high temperatures the anisotropy starts to weaken, and the effective moment approaches its free-ion value $\mu_{\rm eff} = g_J \sqrt{J(J+1)} \approx 3.58 \:\mu_{\rm B}$.

In Figure \ref{fig:Magnetometry}(b) we provide magnetization curves up to fields as large as 40 T. The same easy-axis anisotropy seen in susceptibility remains visible in the entire field range. There is a clear crossover around $\mu_0 H \sim 10$ T, above which the magnetization shows only a weak quasilinear increase. This energy scale roughly matches the hump seen in the effective moment at $T^{*}$ and can be associated with a pseudospin saturation of the lowest few CEF states. Curves collected at 2 K (not shown) and 4 K are nearly identical, indicating that we are below the scale $\theta_{\rm CW}$ where magnetic correlations start to set in.

\subsection{CEF Spectrum}

\begin{figure}[tbp]
\includegraphics[scale=1]{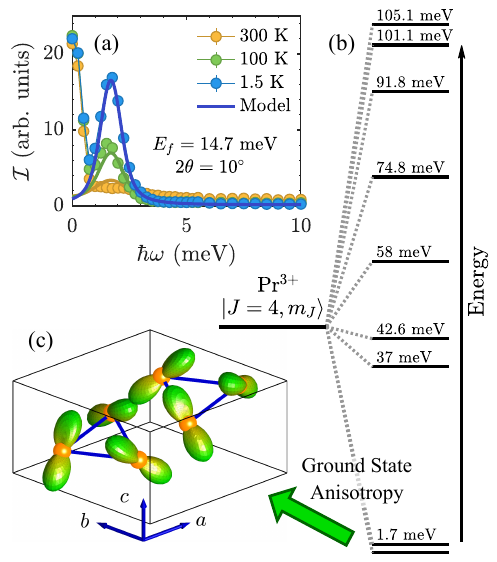}
\caption{(a) Inelastic neutron scattering data on powders reveal a low-lying excitation at $\hbar\omega = 1.69(8)$ meV, its temperature dependence matches expectations for a CEF mode. (b) Sketch of the single-ion crystal field spectrum obtained from the CEF model fit, displaying a well isolated quasi-doublet ground state composed mostly of the Ising states $\ket{J_z = \pm 4}$. (c) A schematic of the magnetic anisotropy for a single unit cell. The surface on each ion represents a projection of the single-ion ground state onto a fully polarized state of arbitrary orientation $| \bra{\psi_{\rm GS}} \ket{J(\phi,\theta)} |^2$, illustrating the "probability" to observe a spin along any given direction.}
\label{fig:EIGER}
\end{figure}

The inelastic neutron spectra measured on $^{11}$B-substituted powders of \pbwo are shown in Figure \ref{fig:EIGER}(a). Apart from the strong quasielastic scattering, the only prominent feature is an intense peak at $\hbar \omega = 1.69(8)$ meV. This mode nicely matches the energy scale seen in magnetometry. It softens slightly with increasing temperature due to thermal expansion and its integrated intensity closely follows the Debye-Waller factor, confirming it is of magnetic origin. Although additional features are observed at higher energies $\hbar \omega > 10$ meV (not shown), they are all shown to be either spurious scattering in the monochromator/analyzer or phonons.

\subsection{Single-Ion Ground State}
\label{sec:GS}

Before attempting to characterize the influence of two-ion interactions, it is key to have a good understanding of the single-ion physics. In rare earths this is facilitated by a clear separation of energy scales \cite{jensenRareEarthMagnetism1991}, allowing us to treat separately the single-ion Hamiltonian
\begin{gather}
	\label{eq:CEF}
    \mathcal{H}_{\rm CEF+Z} = \sum_{n,m} B^m_n \hat{\mathcal{O}}^m_n - \mu_B g_J \mathbf{H} \cdot \hat{\mathbf{J}}
\end{gather}
where $\hat{\mathcal{O}}^m_n$ are the Stevens operators \cite{stevensMatrixElementsOperator1952, hutchingsPointChargeCalculationsEnergy1964} and $g_J$ is the free-ion Landé factor. However, due to the trivial $C_1$ magnetic point symmetry, there are 27 allowed crystal field parameters $B^m_n$, making an unrestricted fit of experimental data to this Hamiltonian severely underconstrained. Instead, we rely on an effective point-charge model \cite{scheiePyCrystalFieldSoftwareCalculation2021, dunEffectivePointchargeAnalysis2021, hutchingsPointChargeCalculationsEnergy1964}: An estimate of the CEF parameters is obtained directly from the experimentally determined crystal structure (see Sec. \ref{sec:Disorder}) by calculating the electrostatic repulsion of neighboring ligands. We can diagonalize the resulting Hamiltonian to obtain the full spectrum and single-ion wavefunctions, allowing a direct comparison with experiment. By varying the effective charges of the three inequivalent O$^{2-}$ positions as well as one mean field exchange parameter $\mathcal{J}_{\rm MF}$ in a simultaneous fit to INS, Magnetization and Susceptibility data, we obtain an accurate approximation of the single-ion Hamiltonian that reproduces all experimental results. 

In Figure \ref{fig:Magnetometry} and Figure \ref{fig:EIGER} the CEF model curves obtained from the effective point-charge fit are compared directly to the data, demonstrating excellent agreement. The CEF parameters and wavefunctions resulting from the fit are provided in Table \ref{tab:CEF_par} and Table \ref{tab:wavefunc}. As expected, the crystal field completely lifts the 9-fold degeneracy. The ground state manifold relevant for low-energy magnetic properties is composed of a low-lying quasi-doublet with a small CEF splitting of $\Delta \approx 1.7$ meV, well isolated from the next excitation around 37 meV. The full CEF spectrum obtained from the fit is displayed in Figure \ref{fig:EIGER}(b). Due to the enormous $\sim 400$ K energy gap above the lowest doublet, the 1.7 meV mode carries nearly all spectral weight at low temperatures, making the higher excitations difficult to observe with INS. Nevertheless, with only four fit parameters the model is fully constrained and consistent with observation. The mean field exchange parameter $\mathcal{J}_{\rm MF} = \frac{1}{2} \sum_{ij} \mathcal{J}_{ij} \sim 0.33$ meV necessary to reproduce the magnetometry data corresponds to a Weiss temperature $\theta_{\rm CW} \approx -1.9$ K, somewhat smaller than suggested by the Curie-Weiss analysis described above.

The local frame of each Pr$^{3+}$ ion is tilted by $\theta \simeq 46.5^{\circ}$ from the crystallographic $c$-axis (see Figure \ref{fig:Structure}(b)). In this reference frame, the quasi-doublet wavefunctions are well approximated by the (anti-)symmetric superpositions
\begin{align}
 	\label{eq:Doublet}
 	\ket{0} &\approx \frac{1}{\sqrt{2}} (\ket{J_z = 4} + \ket{J_z = -4}) \nonumber \\*
    \ket{1} &\approx \frac{1}{\sqrt{2}} (\ket{J_z = 4} - \ket{J_z = -4}).
\end{align}
The single-ion properties are axially symmetric, with perfect Ising anisotropy $g_{zz} \approx 6.09$ and $g_\perp = 0$ due to the non-Kramers nature, while over 95\% of the weights are concentrated on the maximal $\ket{J_z = \pm 4}$ components. The weakened $\sim 2:1$ anisotropy observed in bulk arises naturally from averaging over the 6 equivalent Pr$^{3+}$ ions in the unit cell. A sketch of the ground state anisotropy is depicted in Figure \ref{fig:EIGER}(c), illustrating the probability to observe a spin with any given orientation. Further details about the effective point charge model can be found in Appendix.~\ref{sec:A1}.

\subsection{Model Hamiltonian}
\label{sec:Hamilton}

Due to the large separation of energy scales between crystal field and exchange interactions, only the two lowest levels defined in Eq.~(\ref{eq:Doublet}) are populated at low temperatures and we can construct a low-energy effective Hamiltonian retaining only these quasi-doublet states. The only relevant component from the CEF Hamiltonian will be the splitting $\Delta$ between them. It is well established \cite{wangCollectiveExcitationsMagnetic1968,savaryDisorderInducedQuantumSpin2017,chenIntrinsicTransverseField2019} that an energy splitting between two CEF singlets can be mapped exactly into a transverse magnetic field acting on a corresponding doublet. The resulting single-ion Hamiltonian can be written in terms of local pseudospins $\hat{\mathbf{S}}_i$ as
\begin{gather}
    \mathcal{H} = \Delta \sum_i \hat{S}^x_i - g_{zz} \mu_{\rm B} \sum_i \mathbf{H} \cdot \mathbf{\Hat{z}}_i \hat{S}^z_i,
\end{gather}
where $\mathbf{\Hat{z}}_i$ is the projector onto the local Ising axis. Throughout this paper, the $xyz$-notation always refers to the local basis of the corresponding magnetic ion. Further details about the pseudospin projection are provided in Appendix.~\ref{sec:A2}.

Moreover, the projected two-ion interactions are also severely restricted by the nature of the single-ion ground state. Due to the strict Ising-like anisotropy, the transverse angular momentum components $\hat{J}_\pm$ vanish between the quasi-doublet states. This immediately results in an Ising form in the local basis for $any$ bilinear exchange coupling $\sim \hat{J}_\mu^i K_{\mu \nu}^{ij} \hat{J}_\nu^j$, be it from superexchange or dipolar coupling. Even the higher multipolar interactions typical in rare earths can be ignored here, as they are restricted to rank $\leq 7$ by the maximal angular momentum transferred in each step of the superexchange process and as such cannot connect the leading $\ket{J_z = \pm 4}$ ground state components \cite{rauMagnitudeQuantumEffects2015,iwaharaExchangeInteractionMultiplets2015}.

With this in mind, we write our effective low energy Hamiltonian between local pseudospins $\hat{\mathbf{S}}_i$ as
\begin{gather}
	\label{eq:Hamilton_2lvl}
    \mathcal{H}_{\rm eff} = \Delta \sum_i \hat{S}^x_i + \sum_{ij} \mathcal{J}_{ij} \hat{S}^z_i \hat{S}^z_j - g_{zz} \mu_{\rm B} \sum_i \mathbf{H} \cdot \mathbf{\Hat{z}}_i \hat{S}^z_i
\end{gather}
where $g_{zz} \approx 6.09$ is the longitudinal $g$-factor and the weak dipolar interactions of order $\mathcal{D} = \frac{\mu_0 (\mu_{\rm B} g_J \langle J \rangle)^2}{4 \pi r_{\rm nn}^3} \approx 0.1$ K are simply absorbed into the effective Ising exchange parameters $\mathcal{J}_{ij}$. Potential deviations from this Hamiltonian are constrained by the small $\sim 4\%$ mixing of lower angular momentum states into our refined ground state wavefunctions. As such, we can establish a direct mapping between the low-energy physics of our $J = 4$ Pr$^{3+}$ magnetic moments subject to the crystal electric field and that of Ising-like pseudospin $S = 1/2$ moments in a strong transverse magnetic field - \pbwo turns out to be an excellent realization of the TFIM Hamiltonian. Throughout this paper, we will continue to employ both the $J = 4$ CEF model and the $S = 1/2$ TFIM, pointing out the particular framework used as we go along.

\subsection{Heat Capacity}

\begin{figure}[tbp]
\includegraphics[scale=1]{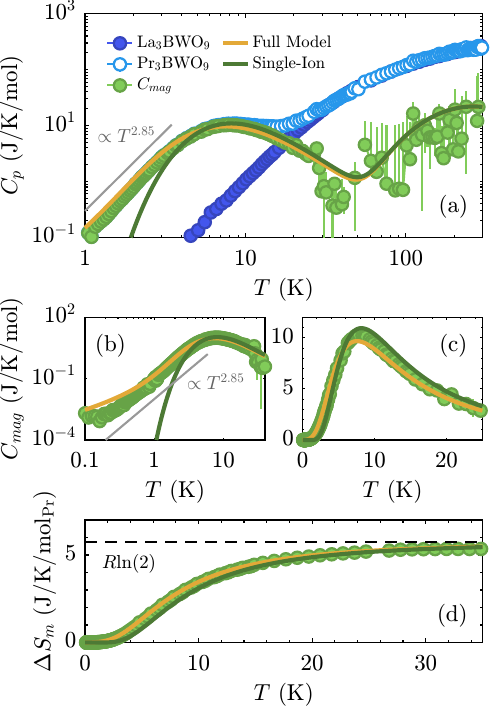}
\caption{(a) Total heat capacity of \pbwo in zero magnetic field. The magnetic contribution is obtained by subtracting a nonmagnetic reference of La$_3$BWO$_9$, shown in more detail on a log-log scale (b) and linear scale (c). The gray solid lines in (a,b) denote the approximate power-law behavior $C_{mag} \propto T^{2.85}$ observed at low temperatures. (d) Change in magnetic entropy calculated through numerical integration of the heat capacity data, pointing to a well isolated doublet ground state. The green solid lines in all panels represent calculated curves for the crystal field model in Sec. \ref{sec:GS}, while the yellow lines include a distribution of crystal field gaps $\mathcal{P}(\Delta)$ expected in presence of structural disorder, as discussed in Sec. \ref{sec:Neutrons}.}
\label{fig:HCZF}
\end{figure}

\begin{figure}[tbp]
\includegraphics[scale=1]{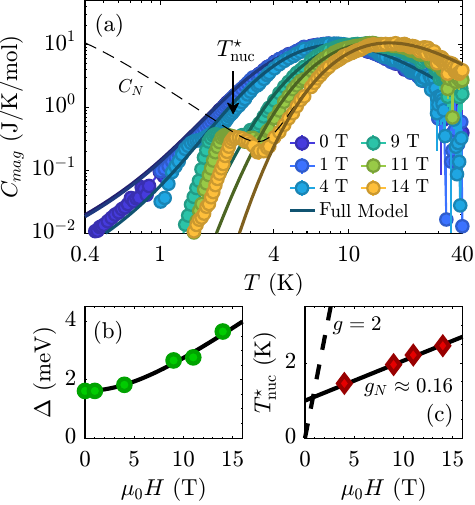}
\caption{(a) Magnetic heat capacity in applied fields along $H \parallel c$. Solid lines represent the calculated magnetic contributions from the crystal field model in Sec. \ref{sec:GS}, including effects of structural disorder. The expected onset of the nuclear contribution at 14 T is marked by a dashed line. (b) Crystal field gap obtained from the Schottky-like maxima, compared to the calculated Zeeman splitting. (c) Field dependence of the peak position associated with the small feature at $T^\star_\mathrm{nuc}$. The slope is much smaller than expected for electronic spins (dashed line), but matches the one calculated for hyperfine-coupled nuclear moments (solid line).}
\label{fig:HCF}
\end{figure}

The zero field heat capacity of \pbwo is shown in Figure \ref{fig:HCZF}, where the magnetic contribution $C_{mag}$ is extracted by subtracting the nonmagnetic La$_3$BWO$_9$ reference data. We observe a broad hump at high temperatures associated with a depopulation of excited crystal field states, as well as another Schottky-like peak centered around 10 K, which matches the energy scale of the $\Delta \approx 1.7$ meV ($\sim 20$ K) level. Comparing these data to the single-ion contribution expected from our CEF model, $C_{mag}$ is nicely reproduced down to $\sim 4$ K, with clear deviations below especially visible on a log scale in Figure \ref{fig:HCZF}(b). A small part of the expected 1.7 meV Schottky feature seems to be shifted towards lower temperatures, where we observe a power-law-like decay $C_p \sim T^\alpha$ with $\alpha \approx 2.85$ instead of the activated behavior of a lone Pr$^{3+}$ ion, with perhaps a tiny upturn below 0.2 K. Importantly there is no sign of long-range order down to $T \lesssim 0.1$ K, consistent with previous reports \cite{zengLocalEvidenceCollective2021}. 
This redistribution of spectral weight could be associated with magnetic correlations, the onset roughly matching the mean field exchange energy $\mathcal{J}_{\rm MF} \sim 0.33$ meV extracted above. Still, the close match to single-ion calculations down to $T \ll \Delta$ might suggest interactions are significantly weaker than the crystal field splitting, making a strongly correlated spin-liquid like state rather unlikely. Although the gapless behavior at low temperatures is seemingly inconsistent with a singlet ground state system, it can arise naturally in the presence of structural disorder as we will show in Sec. \ref{sec:Neutrons}, broadening the mode at $\Delta$ into a distribution of crystal field gaps.
The change in magnetic entropy $\Delta S_m$ seen in Figure \ref{fig:HCZF}(d) is extracted through numerical integration of the heat capacity data. As expected for a well isolated quasi-doublet system, it closely approaches $R \ln(2)$ for temperatures $k_B T \gtrsim \Delta$, justifying the simplified model Hamiltonian derived in Sec. \ref{sec:Hamilton}.

In Figure \ref{fig:HCF}(a) we provide the temperature evolution of the heat capacity in applied magnetic fields along the crystallographic $c$-axis. The large feature associated with the 1.7 meV CEF level moves upwards with field. Its peak position allows us to extract the field dependent CEF gap shown in Figure \ref{fig:HCF}(b). The latter closely follows predictions from our single-ion model, where the zero-field gap is added quadratically to the Zeeman energy as
\begin{gather}
    \Delta (H) = \sqrt{\Delta_0^2 + (g_c \mu_{\rm B} \mu_0 H)^2}.
\end{gather}

Towards lower temperatures an additional broad feature appears at the scale of $T^\star_\mathrm{nuc} \sim 1-2$ K. We estimate its peak position by subtracting the apparent power-law behavior observed at intermediate temperatures, revealing a linear field dependence (see Figure \ref{fig:HCF}(c)). However, the associated $g$-factor is much too small to be of electronic origin, instead matching $g_N \approx 0.158$ \footnote{Given a strong hyperfine coupling, the nuclear spins may slightly polarize the electronic shell, effectively resulting in an enhanced nuclear moment - i.e. an increased nuclear $g$-factor. The enhancement factor $K$ can be calculated explicitly, see Sec.~\ref{sec:muons} for details.} calculated for hyperfine coupled nuclear moments \cite{bleaneyEnhancedNuclearMagnetism1973}. In fact, the onset temperature of this feature agrees nicely with that of the calculated nuclear Schottky contribution, as shown exemplarily for the 14 T data in Figure \ref{fig:HCF}(a). Nevertheless, the magnetic entropy released corresponds only to $\sim5\%$ of $ R \ln(2)$ and the feature becomes suppressed slightly below the onset temperature of $C_N$. 

Although a drastically reduced nuclear heat capacity might seem puzzling, there are several potential mechanisms for bringing this about. For example, it is known that electronic spin fluctuations can lead to a reduction in the nuclear Schottky contribution \cite{bertinEffectiveHyperfineTemperature2002}. The corresponding fluctuations in the hyperfine field can diminish the population differences between the $^{141}$Pr Zeeman levels, resulting in a suppressed $C_N$, seen e.g. in Pr-based pyrochlores \cite{maclaughlinUnstableSpiniceOrder2015, martinDisorderQuantumSpin2017a}. 
Another possibility would be a partial decoupling of the nuclear degrees of freedom due to a lack of phonons (or magnons) at these temperatures \cite{abragamPrinciplesNuclearMagnetism,isonoSpinlatticeDecouplingTriangularlattice2018a}. The nuclear spin system only couples indirectly to the external heat bath through excitations of the lattice/electronic spins, so without an abundance of such low energy modes we cannot probe efficiently the nuclear heat capacity. A similar decoupling phenomenon was observed in the sister material \nbwo above the pseudospin saturation \cite{flavianMagneticPhaseDiagram2023a}.

In \pbwo, the purported nuclear contribution is suppressed immediately below the onset temperature of the calculated $C_N$, which heavily favours the latter scenario. For fast fluctuating electronic spins, the peak position should be preserved, while a decoupling of nuclear spins would become effective for $C_N \gg C_{lat}, C_{mag}$, meaning that only the high temperature tail of the nuclear heat capacity can be observed. Towards lower fields, the hyperfine splitting decreases significantly, further suppressing $C_N$ until it is no longer visible below $\mu_0 H \simeq 1$ T.

\subsection{Ultrasound}

\begin{figure}[tbp]
	\centering
	\includegraphics[scale=1]{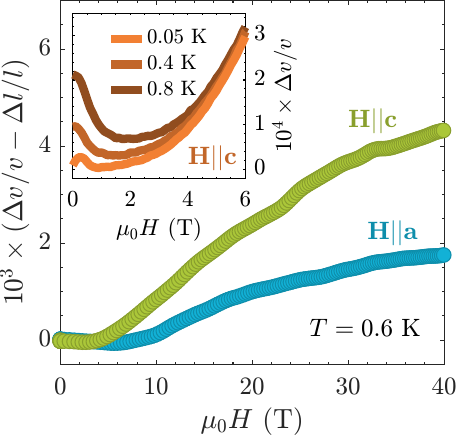}
	\caption{Field dependence of the relative change in sound velocity of \pbwo, measured along several crystallographic axes. The temperature dependence at low fields is depicted in the inset. In principle, the experiment is sensitive also to changes in sample length, though this contribution is generally negligible. The small wiggles at high fields are caused by noise associated with the rapid (dis-)charging of the magnet, not to be confused with real features in the data.}
	\label{fig:Sound}
\end{figure}

Given the strong spin-orbit coupling inherent in rare earths, it is instructive to consider the effects of magnetoelastic coupling, as even small changes in the crystal structure can potentially have drastic consequences on the magnetism (and vice versa). To that end, we probe the ultrasound velocity in \pbwo, focusing on the longitudinal $c_{33}$ mode propagating with $v_s \simeq 3.3(1)$ km/s. Figure \ref{fig:Sound} displays the relative change in sound velocity $\Delta v/v$ as a function of magnetic field. In agreement with the heat capacity data, we do not detect any signs of phase transitions down to $T \lesssim 600$ mK even for fields exceeding 40 T, suggesting that \pbwo retains a quantum disordered ground state in the entire $H-T$ phase space. Instead, we observe a small decrease in sound velocity at low fields (i.e. a small "softening" of the lattice), followed by a strong quasi-linear increase. Both the earlier crossover between these two regimes and the overall stronger magnetoelastic response for fields along $H \parallel c$ are linked to the easy-axis nature of \pbwo, explained mostly by a difference in $g$-factors. Furthermore, $\Delta v/v$ qualitatively follows the single magnon gap $\Delta (H)$ (see Figure \ref{fig:HCF}(b)), where the crossover appears when the crystal field splitting $\Delta$ acts on the same scale as the Zeeman energy $g \mu_{\rm B} \mu_0 H$. 
In fact, these observations are not surprising, as the dominant contribution to the magnetoelastic energy density in rare earths typically originates from a linear single-ion strain interaction \cite{luthiPhysicalAcousticsSolid1967}. Lattice vibrations modulate the crystal electric field, which directly determines the orbital state of each ion, resulting in a magnetoelastic response dictated by single-ion properties. Though the overall scale of $\Delta v/v$ of course depends on the field orientation, the sign of the effect does not. Because there is no transverse $g$-factor, fields can only couple to the local $\hat{S}^z$ pseudospin component, resulting in an analogous deformation to the $4f$ charge density independent of field direction. 
We note that in principle, the phase sensitive detection method utilized here responds also to magnetostrictive changes in the sample length $\Delta l/l$ along the propagation direction, i.e. we measure the phase shift $\Delta \phi/\phi = \Delta l/l - \Delta v/v$ as denoted in Figure \ref{fig:Sound}. However, barring any (magneto-)structural phase transitions or incipient soft modes thereof, the change in lattice spacing should be orders of magnitude weaker than the present effects and will therefore be ignored here.

Perhaps more interesting is the significant temperature dependence shown in the inset of Figure \ref{fig:Sound}. At zero field, there is a rapid increase in sound velocity upon heating, which becomes heavily suppressed at fields as low as $\mu_0 H \lesssim 0.5-1$ T, gradually giving way to $T$-independent behavior. All data are taken below $k_B T \lesssim \Delta/20$, so effects of CEF population should be negligible and these features more likely result from a change in the spin correlations.
But even though such a lattice "hardening" at zero field can be easily rationalized as a weakening of magnetic correlations, it is unclear how its sudden suppression at extremely low fields could be explained in this scenario. Indeed, there is no obvious energy scale associated with this suppression - both Weiss temperature and crystal field gap are significantly larger - and neither susceptibility nor magnetization exhibit significant features at such low fields, making this behavior somewhat anomalous. 

All in all, magnetoelastic effects appear to be rather subdued in \pbwo, inducing only gradual changes in sound velocity mostly accounted for by single-ion effects, seemingly proving unimportant to our understanding of the minimal magnetic model.

\subsection{Muon Spin Relaxation}
\label{sec:muons}

\begin{figure}[tbp]
	\includegraphics[scale=1]{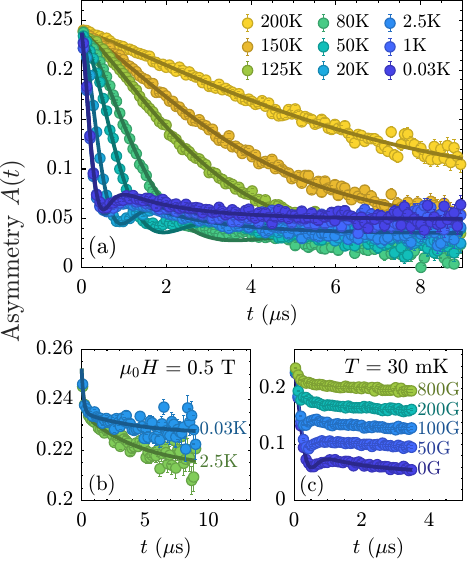}
	\caption{(a) Zero-field $\mu$SR asymmetry spectra at several temperatures, along with KT-like fits discussed in the main text. (b) The asymmetry at high fields decays exponentially, though a small step in the non-relaxing fraction $G_0$ is seen around 2 K, just as in zero field. (c) Field dependence of the $\mu$SR relaxation at base temperature, decoupling the small quasi-static moments.}
	\label{fig:MuonData}
\end{figure}

\begin{figure}[tbp]
	\includegraphics[scale=1]{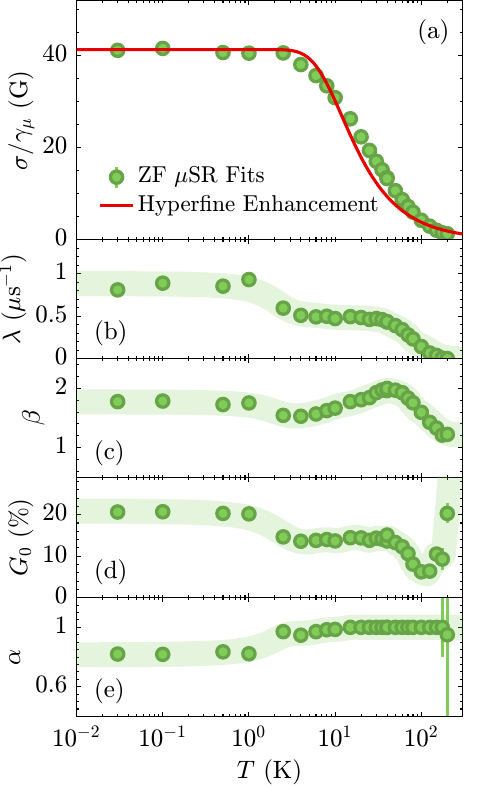}
	\caption{Temperature dependence of the zero field $\mu$SR fit parameters based on Eq.~(\ref{eq:musr}). The red line in (a) indicates the calculated temperature dependence in a hyperfine enhancement scenario, where the gap $\Delta$ is taken from our crystal field model. The green shading in the other panels represents a guide to the eye.}
	\label{fig:MuonFits}
\end{figure}

As a complement to thermodynamic probes, the ground state magnetism in \pbwo is characterized by means of muon spin relaxation. The asymmetry spectra of the zero-field $\mu$SR data at selected temperatures are displayed in Figure \ref{fig:MuonData}(a). Contrary to expectations, the low temperature data show a clear dip at $t \sim 0.5$ $\mu$s followed by the partial recovery of a weakly decaying tail, amounting to $1/3$ of the initial polarization. The signal remains temperature independent below $T \sim 2$ K, whereas at higher temperatures the dip weakens and shifts to later times, gradually giving way to paramagnetic relaxation above $T \gtrsim 100$ K. A Kubo-Toyabe-like (KT) $1/3$-tail provides strong evidence for quasi-static magnetism at the $\mu$SR timescale \cite{leeMuonScienceMuons2017}, whereas the lack of further oscillations implies a frozen disordered state instead of collective long-range order.

In longitudinal fields of order $\mu_0 H \lesssim 0.1$ T, the KT-relaxation becomes fully suppressed, leaving only the double-exponential behavior seen in Figure \ref{fig:MuonData}(c). This clearly affirms the dynamical nature of the weakly relaxing tail, but also proves that the quasi-static internal fields responsible for the zero field relaxation must be relatively weak. Indeed, the width of the field distribution in a KT-like model can be estimated from the dip position $ t_{\mathrm{Dip}}$ as $\sigma/\gamma_\mu \sim 2/(\gamma_\mu t_{\mathrm{Dip}}) \sim 4$ mT, where $\gamma_\mu$ refers to the muon gyromagnetic ratio.

We model the observed zero field depolarization using the following relaxation function
\begin{gather}
	G(t) = G_{\rm VKT}(t) \exp(-(\lambda t)^\alpha) + G_0
	\label{eq:musr}
\end{gather}
where
\begin{gather}
	G_{\rm VKT}(t) = \frac{1}{3} + \frac{2}{3}(1 - (\sigma t)^\beta)\exp(-(\sigma t)^\beta/\beta)
\end{gather}
corresponds to a Voigtian Kubo-Toyabe profile \cite{crookVoigtianKuboToyabe1997}, appropriate for random static moments with a field distribution interpolating between the Gaussian ($\beta = 2$) and Lorentzian ($\beta = 1$) case. The stretched exponential accounts for the weak dynamical relaxation observed at late times. A stretching exponent $\alpha$ is often introduced phenomenologically to account for the presence of several different relaxation times \cite{johnstonPRB742006}, be it from spatial inhomogeneity in glassy systems \cite{leeMuonScienceMuons2017} or simply due to a handful of stopping sites \cite{dingPRB1022020}. Finally, $G_0$ represents a constant asymmetry fraction within our time window, including a background contribution from muons stopping outside the sample.

With this model, we can successfully account for the zero field relaxation in the entire temperature range 30 mK $\lesssim T \lesssim$ 200 K - the resulting fit parameters are provided in Figure \ref{fig:MuonFits}. 
The width of the internal field distribution $\sigma/\gamma_\mu$ remains constant below 5 K before smoothly starting to decrease, only tending towards zero at the highest measured temperatures.
Both the relaxation rate $\lambda$ and the KT-exponent $\beta$ show a broad hump around 50 K, consistent with a slow crossover towards paramagnetic relaxation at high temperatures. 
Apart of that, the only prominent feature in the data is a step-like change seen around $T \sim 2$ K, present in all fit parameters except $\sigma/\gamma_\mu$. 
Especially the step in $G_0$ is directly visible in Figure \ref{fig:MuonData}(a) as a shift of the late-time asymmetry tail between the 1 K and 2.5 K curves. 
In fact, the same feature remains present in a $\mu_0 H = 0.5$ T longitudinal field (see Figure \ref{fig:MuonData}(b)), even though the weak internal fields must be fully decoupled in this regime, suggesting a dynamical origin for the anomaly.
We can explain this behavior by invoking the existence of several stopping sites \footnote{In oxides, the muon tends to stop $\sim 1\text{ \AA}$ away from a nearby oxygen site, analogous to an OH bond \cite{leeMuonScienceMuons2017}. Due to the complex low-symmetry crystal structure with 18 oxygen ions per unit cell, careful DFT calculations would be required to determine the potential candidate stopping sites. Such an analysis is beyond the scope of this work - it would not help in analyzing our $\mu$SR results, as any model analogous to Eq.~(\ref{eq:musr}) with several stopping sites strongly over-parameterizes the data.}: The residual dynamics in some fraction of the sample slows down at low temperatures until the tail relaxation of that component approaches zero within our time window. This brings about an increase in the "background" fraction $G_0$, as well as a change in the effective relaxation rate $\lambda$. It is also consistent with the appearance of a stretching exponent $\alpha < 1$ for the dynamical relaxation, which is expected in the presence of several different relaxation times. The only magnetic energy scale roughly matching this $T \sim 2$ K feature would be the that of the two-ion interactions $\mathcal{J}_{\rm MF} \sim 0.33$ meV, indicating that it could be related to the onset of some residual AFM correlations, which quench/slow down the dynamics in a fraction of the sample.

In order to pinpoint the nature of the weak quasi-static magnetism, we compare the $\sigma/\gamma_\mu \simeq 4$ mT observed at low temperatures to typical internal field values expected in different scenarios. For instance, similar Pr-based materials that order magnetically typically exhibit a strong $\sim 0.2 - 0.5$ T dipolar field at the muon site \cite{AnandPRB1072023}, roughly two orders of magnitude larger than seen in experiment. This clearly contradicts the possibility of a glassy ground state, where the $\mu$SR relaxation would be dominated by fully frozen electronic moments $\mu_J = 3.58 \:\mu_{\rm B}$. Moreover, the internal field distribution is also atypical of nuclear magnetism (the $^{141}$Pr nuclei carry a spin $I = 5/2$), which usually leads to a temperature-independent width in the range of $\sim 0.1$ mT - seemingly ruling out nuclear moments as a source of the KT-like behavior. But as we are dealing with a non-Kramers system endowed with strong hyperfine coupling and a low-lying CEF level, there is a well known mechanism for amplifying the effective nuclear moments, aptly named enhanced nuclear magnetism \cite{bleaneyEnhancedNuclearMagnetism1973,blundellQuantumMuon2023}. Due to the large Van-Vleck susceptibility, the nuclear spins can partially polarize the electronic shell, giving rise to "dressed" nuclear moments
\begin{gather}
	m_z^N = g_I \mu_{\rm B} I_z (1 + K),
\end{gather}
effectively enhanced by a factor $1+K$ compared to the bare nuclei. Enhancement factors of order $10-100$ are commonly observed in Pr-based magnets \cite{zorkoGroundStateEasyAxis2010, maclaughlinUnstableSpiniceOrder2015,shuMuonSpinRelaxation2007}, providing a natural explanation for the data in terms of hyperfine enhanced randomly oriented nuclear spins, with perhaps some slow persistent fluctuations at a rate of $\sim 1$ MHz.

We inspect the validity of this hyperfine enhancement scenario by calculating the temperature dependence of the effective nuclear moments, which should scale directly with the width of the field distribution $\sigma/\gamma_\mu \propto m_z^N$. For a two-level system \cite{bleaneyEnhancedNuclearMagnetism1973}, we find
\begin{gather}
	K = \frac{2 g_J A_J \bra{1} J_z \ket{0}^2}{g_I \Delta} \tanh \left(\frac{\Delta}{2 k_B T} \right),
\end{gather}
where the hyperfine coupling constant $A_J = 0.05$ K \cite{kondoInternalMagneticField1961a}, the matrix element $\bra{1} J_z \ket{0} \approx 3.81$, as well as the $g$-factors and crystal field gap $\Delta$ are all fully determined, leaving no free parameters on $K$ and resulting in an enhancement factor as large as $K \simeq 70$ at low temperatures. In Figure \ref{fig:MuonFits}(a) we compare the calculated temperature dependence of $m_z^N$ to the observed field distribution $\sigma/\gamma_\mu$, giving excellent qualitative agreement. Towards higher temperatures, excited CEF levels gradually become populated, which quenches our enhancement mechanism and begets a drop in $\sigma/\gamma_\mu$. The small residual differences at intermediate temperatures are unsurprising, considering that we have ignored the potential influence of muon-induced distortions, which have been shown to significantly modify the CEF spectrum in non-Kramers ions, resulting in a range of gaps $\Delta$ depending on the distance from the stopping site \cite{forondaAnisotropicLocalModification2015}.

\subsection{Inelastic Neutron Scattering}
\label{sec:Neutrons}

\begin{table}
	\renewcommand{\arraystretch}{1.2}
	\caption{Model parameters of the minimal magnetic Hamiltonian for \pbwo, obtained from a LSWT analysis of inelastic neutron data. We can convert between the $S=1/2$ TFIM parameters $\mathcal{J}_{ij}$ (center column) and the $J=4$ CEF model parameters $\tilde{\mathcal{J}_{ij}}$ (right column) by using Eq.~\ref{eq:ConvertJ}.}
	\centering
	\begin{tabularx}{0.45\textwidth}{c Y Y Y}
		\toprule\toprule
		&  Distance (\AA) & Value (meV) $S=1/2$ & Value (meV) $J=4$ \\
		\midrule
		$\Delta$ & - & 1.62(1) & 1.62(1) \\
		$\mathcal{J}_1$ & 3.96 & 0.30(1) & 0.0068(2) \\
		$\mathcal{J}_1'$ & 3.96 & -0.11(1) & -0.0026(2) \\
		$\mathcal{J}_\Delta$ & 4.26 & 0.06(1) & 0.0053(8) \\
		\bottomrule
	\end{tabularx}
	\label{table:Neutron}
\end{table}

\begin{figure*}[tbp]
	\centering
	\includegraphics[scale=1]{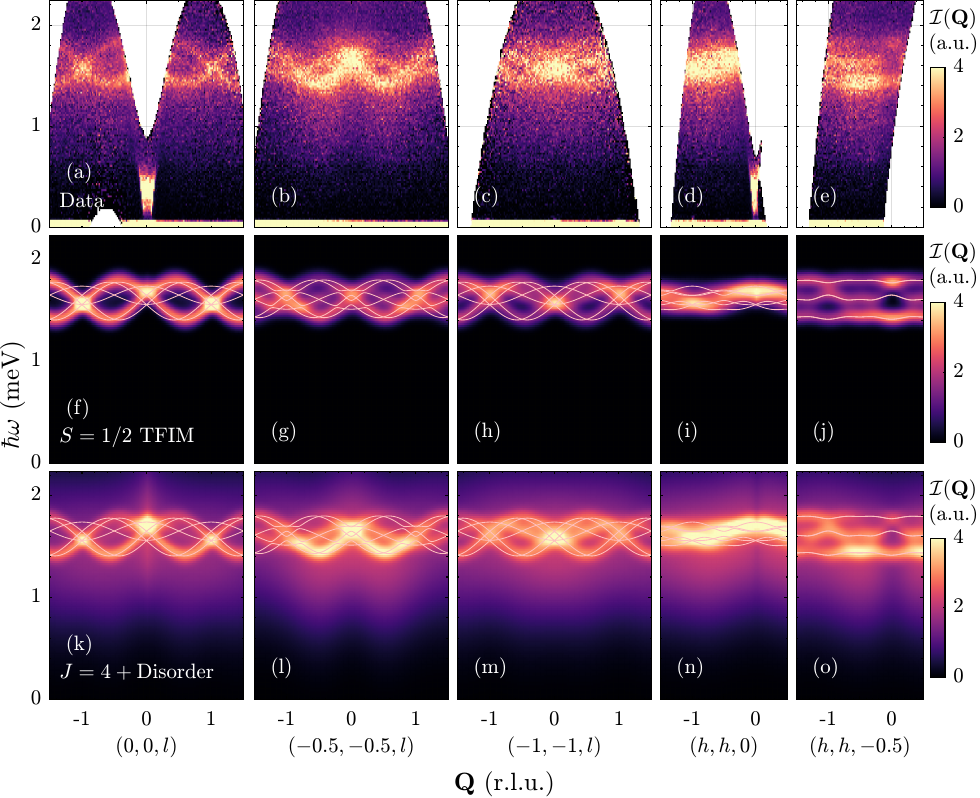}
	\caption{(a-e) Inelastic neutron scattering spectra in \pbwo taken at $T = 0.3$ K. Intensities are integrated by $|\mathbf{Q}| \leq 0.15$ r.l.u. in momentum transfer perpendicular to the plot axis. (f-j) LSWT calculations for the $S = 1/2$ transverse field Ising model with parameters from Table \ref{table:Neutron}. The sharp modes are reproduced, whereas the continuum is not. (k-o) $SU(9)$ spin wave calculations for the full $J = 4$ model including a phenomenological treatment of the structural disorder, fully reproducing the excitation spectrum.}
	\label{fig:NeutronQE}
\end{figure*}

\begin{figure*}[tbp]
\centering
\includegraphics[scale=1]{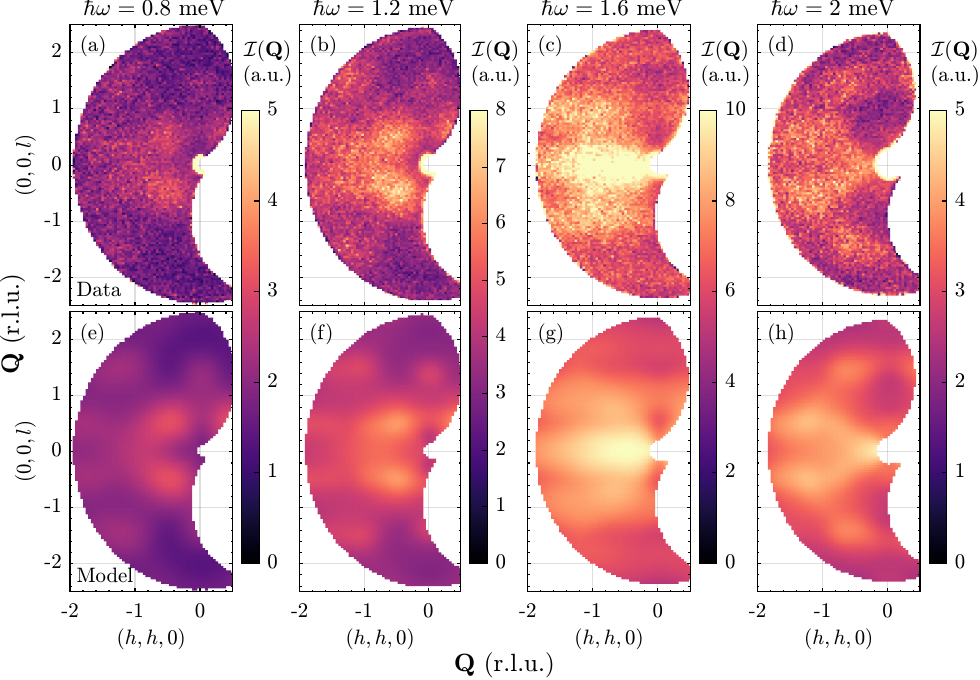}
\caption{(a-d) Constant-energy slices through the INS spectra, integrated by $|\hbar\omega| \leq 0.2$ meV in energy and $|\mathbf{Q}| \leq 0.15$ r.l.u. in momentum transfer perpendicular to the scattering plane. (e-h) $SU(9)$ spin wave calculations for the full $J = 4$ model with parameters listed in Table \ref{table:Neutron}, reproducing the excitation spectrum.}
\label{fig:NeutronQQ}
\end{figure*}

The inelastic neutron spectra collected at AMATERAS are summarized in Figure \ref{fig:NeutronQE} and Figure \ref{fig:NeutronQQ}. These are raw data without background subtraction, taken at $T \approx 300$ mK. The spectrum  in \pbwo seems to be composed of two kinds of excitations, featuring a coexistence of sharp spin-wave-like modes and broad continuum excitations.

The former are composed of at least three separate bands centered at $\hbar\omega \sim 1.6$ meV, matching the energy scale of the CEF gap $\Delta$. We assign them as dispersive crystal field excitons, local excitations hopping between sites due to the presence of two-ion interactions. The associated bandwidth $2\mathcal{J}_{\rm MF} \sim 0.6$ meV is fully consistent with our mean field estimate $\mathcal{J}_{\rm MF} \sim 0.33$ meV obtained from the crystal field model. Although there is clearly both dispersion and $Q$-modulation present within the kagome planes, the dominant dispersion direction seems to point out-of-plane along the $c$-axis, confirming the importance of the nearest-neighbor couplings $\mathcal{J}_1$ and $\mathcal{J}_1'$.

On the other hand, the continuum scattering is also centered close to the CEF energy scale $\Delta$ and can be described by a broad power-law decay towards low energies, with a full width at half maximum around 1.2 meV. It accounts for most of the inelastic scattering intensity, responsible for roughly $80 \%$ of the total spectral weight. Its momentum dependence seems to partially mimic that of the exciton modes.

In order to analyze the observed spectrum, we fit a number of constant-$Q$ cuts to a combination of one Gaussian-broadened (continuum) and several resolution-limited (spin waves) modes, extracting almost 600 dispersion points across the entire Brillouin zone. These data are analyzed by means of linear spin-wave theory (LSWT) calculations using the SUNNY software package \cite{sunny_git}. We perform a global LSWT fit to the observed mode energies, based on the two-level TFIM Hamiltonian Eq.~(\ref{eq:Hamilton_2lvl}) derived above. In accordance with our thermodynamic data, the ground state is assumed to be quantum disordered in these calculations, corresponding to a fully polarized state in the pseudospin language of our transverse-field Ising model, where each spin points along its local $\mathbf{\hat{x}}$-axis. Therefore, LSWT is expected to be an excellent approximation and should allow for an accurate estimation of the "true" (non-renormalized) exchange constants \cite{coldeaDirectMeasurementSpin2002}.

The optimized model parameters from this analysis are listed in Table \ref{table:Neutron} and the resulting simulated spectra, including magnetic form factor corrections, are displayed in Figure \ref{fig:NeutronQE}(f,g,h,i,j). Besides the gap $\Delta$, we require three exchange parameters to account for the experimental dispersion: The nearest-neighbor coupling $\mathcal{J}_1$ is responsible for the bandwidth along $l$, while a smaller $\mathcal{J}_1'$ with opposite sign is necessary to reproduce the nodes at integer $l$-values. Finally, $\mathcal{J}_\Delta$ tunes the offset of the two flatter bands, as well as the small in-plane dispersion. All the above parameters are well decoupled and fully constrained by the fit. Attempts were made to introduce additional interactions $\mathcal{J}_\nabla$ and $\mathcal{J}_3$, but this does not improve the agreement with experiment and the exchange constants from such a fit remain zero within errors ($\lesssim 0.02$ meV). We can accurately reproduce all experimentally observed modes with this model, accounting for all six expected spin-wave branches with good qualitative agreement on the intensity distribution. Nevertheless, the broad continuum is missing entirely in our model.

The calculated intensities can be improved significantly by using the full $J = 4$ CEF Hamiltonian
\begin{gather}
	\label{eq:H_full}
	\mathcal{H}_{\rm full} = \sum_{n,m} B_n^m \hat{\mathcal{O}}_n^m + \sum_{ij} \tilde{\mathcal{J}_{ij}} \hat{\mathbf{J}}_i \cdot \hat{\mathbf{J}}_j - \mu_{\rm B} g_J \sum_i \mathbf{H} \cdot \hat{\mathbf{J}}_i,
\end{gather}
i.e. working with $SU(9)$ coherent states \cite{dahlbomRenormalizedClassicalTheory2023} instead of $S = 1/2$ dipoles polarized along the transverse field direction. Due to the mapping between the $S=1/2$ TFIM and $J=4$ CEF model, the exchange parameters $\tilde{\mathcal{J}_{ij}}$ in the latter framework can be obtained directly by rescaling those from the LSWT fit as
\begin{gather}
	\label{eq:ConvertJ}
	\tilde{\mathcal{J}_{ij}} = \frac{1}{\mathbf{\Hat{z}}_i \cdot \mathbf{\Hat{z}}_j} \left( \frac{1/2}{\bra{1} J_z \ket{0}} \right)^2 \mathcal{J}_{ij},
\end{gather}
where the matrix element $\bra{1} J_z \ket{0} \approx 3.81$ accounts for the change in spin length. The geometrical factor $\mathbf{\Hat{z}}_i \cdot \mathbf{\Hat{z}}_j$ reflects the fact that we are using Heisenberg couplings in the CEF model; projecting down to the low-energy degrees of freedom will anyways render void all exchange components not along the local Ising axes. We note that because of the much higher energy resolution on AMATERAS compared to the CEF spectra obtained on EIGER, the crystal field parameters in Table \ref{tab:CEF_par} are also rescaled slightly by a global factor $f_\Delta = \frac{\Delta_{\mathrm{AMATERAS}}}{\Delta_{\mathrm{CEF-fit}}} \approx 0.96$ to obtain satisfactory agreement. This allows us to fine tune the observed gap energy without changing the anisotropies refined in bulk. The appropriately rescaled parameters for the $J = 4$ CEF model are listed in Table \ref{table:Neutron}. They produce the exact same dispersion as the TFIM couplings, but give a quantitative agreement on the relative intensity distribution. The enhanced $\tilde{\mathcal{J}}_\Delta / \tilde{\mathcal{J}}_1$ ratio in this framework simply reflects the difference in angle $\cos \Omega_{ij} = \mathbf{\Hat{z}}_i \cdot \mathbf{\Hat{z}}_j$ between the easy axes of first and second neighbors.

Finally, we attempt to empirically model the broad continuum excitations. Since all spin-wave branches are accounted for across the Brillouin zone and the thermodynamic probes rule out the presence of further accidentally degenerate CEF modes, it is unclear how any scenario of fractionalized excitations could give rise to the continuum seen in \pbwo. Instead, the fact that the latter seems to mimic the spin-wave dispersion favors a scenario of structural disorder. Due to the lack of time reversal symmetry in Pr$^{3+}$, the splitting $\Delta$ between our quasi-doublet states is highly sensitive to changes in the local crystal structure \cite{martinDisorderQuantumSpin2017a, liSpinDynamicsGriffiths2021}. As such, any local distortion due to e.g. strains, impurities or domain walls will slightly alter the CEF scheme, resulting in a distribution of gaps $\mathcal{P}(\Delta)$. As we observe both broad and sharp contributions in our spectra, we have to assume that some ions are sufficiently removed from such impurity sites to give rise to the resolution limited spin-waves, while the remaining ones possess a random distribution of gaps, accounting for the continuum. An investigation of the local strain fields in \pbwo goes beyond the scope of this work, though we find that a simple Gaussian distribution $\mathcal{P}(\Delta)$ is sufficient for modeling the spectrum.

Indeed, we can successfully reproduce our inelastic data in this disorder scenario by adding a Gaussian-broadened copy of the calculated spin-wave spectrum with the appropriate spectral weight. To that end, we introduce three new global parameters: The fraction $f_B \approx 85\%$ of affected magnetic sites, the Gaussian width $\sigma_B \approx 0.5$ meV reflecting the degree of randomness and the mean $\Delta_B \approx 1.51$ meV of the distribution (to obtain complete agreement with experiment we choose $\Delta_B \neq \Delta$). Presumably, quenched disorder would also induce some variation in the exchanges, but here we push all randomness onto the Gaussian width $\sigma_B$, in order not to over-parameterize our model. The calculated spectra obtained with these empirically optimized parameters in the $J = 4$ CEF model framework are displayed in Figures \ref{fig:NeutronQE}(k,l,m,n,o) and \ref{fig:NeutronQQ}(e,f,g,h). Our model can semi-quantitatively account for all experimental features, including the coexistence of resolution-limited and broadened modes, as well as the intensity distribution across the entire Brillouin zone.

As a cross check, we calculate the single-ion heat capacity in the proposed disorder scenario, shown in Figure \ref{fig:HCZF} as composed of Schottky terms with the distribution of gaps $\mathcal{P}(\Delta)$ extracted from the neutron data. Unlike the "clean" single-ion calculation, this model closely approximates the measured $C_p(T)$ curve in the entire temperature range with no free parameters, including the power-law like behavior at low temperatures. As seen in Figure \ref{fig:HCF}, we can even reproduce the results at finite fields, although the small hump seen at low temperatures remains absent, consistent with our explanation in terms of partially decoupled nuclear degrees of freedom. This striking parameter-free agreement, together with the excellent fit of our inelastic neutron data, strongly justify our model assumptions, confirming that the magnetism of \pbwo should be understood in terms of dispersive crystal field excitons, with many sites affected by some form of structural disorder that locally tunes the average CEF scheme.

\subsection{Crystal Structure \& Disorder}
\label{sec:Disorder}

\begin{figure}[tbp]
	\includegraphics[scale=1]{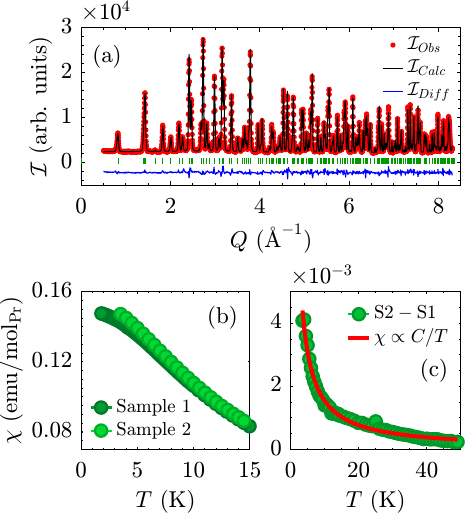}
	\caption{(a) Rietveld refinement of neutron powder diffraction pattern obtained at $T = 1.5$ K in \pebwo. (b) Magnetic susceptibility curves collected on several single crystal samples. (c) The difference in $\chi$ between measurements on samples S1 and S2 can be fit to a Curie tail, consistent with a small $\sim 1.1\%$ concentration of Pr$^{4+}$ impurities.}
	\label{fig:Disorder}
\end{figure}

Given the success of our disorder model in explaining both the thermodynamics and spectrum of \pbwo, it is well worth delving into the potential causes for disorder by taking a closer look at the crystal structure. To that end, we collect neutron powder diffraction data on $^{11}$B-substituted samples and carry out Rietveld refinements, the results of which are shown in Figure \ref{fig:Disorder}(a). The final $R$-factor including anisotropic thermal parameters amounts to $R_p = 7.5 \%$ and the refined atomic positions (utilized for the point-charge calculations in Sec. \ref{sec:GS}) are listed in Table \ref{tab:Structure}. 

Perhaps the most common source of structural disorder found in quantum magnets is antisite mixing, a random switching of ions with similar atomic radii which is notorious for its role in spin-liquid candidates such as Herbertsmithite \cite{hanCorrelatedImpuritiesIntrinsic2016} or YbMgGaO$_4$ \cite{liCrystallineElectricFieldRandomness2017, zhuDisorderInducedMimicrySpin2017}. The large differences between atomic radii makes this a rather unlikely problem in \pbwo \cite{ashtarNewFamilyDisorderFree2020}, which is confirmed by our refinement attempts, showing no signs of antisite disorder within uncertainties ($1-2$\%).

Another common concern is off-stoichiometry, as seen e.g. in pyrochlore oxides \cite{maclaughlinUnstableSpiniceOrder2015, koohpayehSynthesisFloatingZone2014}. A refinement with variable site occupancy indeed points to a slight excess in oxygen concentration of $\sim 6(2)\%$ compared to the other ions, yielding an $0.2\%$ improvement in $R$-factors. In Pr-based oxides, the charge imbalance created by the extra O$^{2-}$ is typically accommodated by increasing the valence for some fraction of magnetic ions from Pr$^{3+}$ to Pr$^{4+}$ \cite{koohpayehSynthesisFloatingZone2014, ferroPhysicochemicalElectricalProperties2011}. Even though all powders were sintered for several days at 1200$^{\circ}$C during synthesis until a uniform green color was obtained (indicative of pure Pr$^{3+}$ valence \cite{ferroPhysicochemicalElectricalProperties2011}) and the XRD spectra no longer changed, it would seem that some trace amounts of Pr$^{4+}$ and O$^{2-}$ impurities have remained throughout this process.

If our samples really are contaminated with magnetic Pr$^{4+}$ impurities, these should give rise to an observable Curie-like tail in the low temperature susceptibility. To test this hypothesis, we compare magnetic susceptibility curves collected on several different single crystals. Subtracting $\chi(T)$ from two samples grown from different batches as depicted in Figure \ref{fig:Disorder}(b,c), the difference can indeed be fit to a $\chi \propto 1/T$ contribution, amounting roughly to a $\sim 1.1 \%$ concentration of impurity spins. Although this number corresponds to the {\it difference} between Pr$^{4+}$ concentrations from two separate samples, we can also establish an upper bound $\lesssim 2\%$ on the {\it absolute} number of impurity spins. This is because the Van-Vleck susceptibility of a singlet ground state system for $T \rightarrow 0$ is constant - i.e. the susceptibility should not have a positive slope after subtracting a potential Curie-like impurity tail. A significantly smaller concentration than that obtained from powder diffraction is to be expected, as single crystal samples are commonly of higher quality than powders.

With this, we have confirmation from two separate sources - neutron powder diffraction and magnetic susceptibility measurements - that a small surplus of oxygen and the concomitant Pr$^{4+}$ charge impurities are responsible for the structural disorder detected through other probes. In elasticity theory, the strain fields produced by such point defects result in lattice distortions that decay as a power law $\sim 1/r^2$ with distance \cite{TeodosiuElasticPointDefect1982} – analogous to the underlying Coulomb forces at the root of these displacements. While the resulting change in CEF parameters is difficult to estimate – requiring detailed knowledge on the atomic displacements – impurities should clearly have {\it long-ranged} effects. So although $\lesssim 2 \%$ might seem like a small amount of disorder, the vast majority of sites will be "affected" by the long-ranged structural distortions emanating from the local impurity sites, creating small modifications to the local crystal electric field. Such corrections would be irrelevant for most common magnets, but the lack of Kramers degeneracy makes \pbwo particularly susceptible to this effect. As such, the seemingly huge fraction $f_B \approx 85 \%$ of sites contributing to the broad continuum scattering discussed in Sec. \ref{sec:Neutrons} can be naturally explained by a small number of impurities $\lesssim 2\%$.

\section{Discussion}

While \pbwo was initially advertised as a proximate spin-liquid candidate on the breathing kagome lattice \cite{ashtarNewFamilyDisorderFree2020,zengLocalEvidenceCollective2021}, a more careful investigation reveals a completely disparate picture. We combine thermodynamic, magnetometric, neutron- and muon spectroscopy data with numerical modeling, providing strong evidence that \pbwo can be understood as a singlet ground state system with a small gap $\Delta = 1.62$ meV separating the lowest two crystal field levels. After a pseudospin projection down to this lowest quasi-doublet, the effective $S = 1/2$ Hamiltonian takes the form of a nearly ideal Ising model in a transverse magnetic field. Despite the trivial point symmetry and complicated structure, the effects of crystal field and non-Kramers moments with large-$J$ conspire to form a simple model Hamiltonian of significant interest.

The phase diagram was thoroughly characterized, proving that \pbwo remains disordered in the entire $H-T$ phase space. Although both the triangle-based lattice topology and the FM-AFM exchange competition contribute to a significant magnetic frustration, neither is the main cause of the quantum disordered ground state. Even if we completely disregard frustration effects, the exchange interactions barely amount to $30 \%$ of the critical ratio 
\begin{gather}
	A_{cr} = \frac{2 \bra{1} J_z \ket{0}^2 \mathcal{J}(\mathbf{Q})}{\Delta} = 1
\end{gather}
necessary for induced-moment order on a mean-field level \cite{thalmeierInducedQuantumMagnetism2023,jensenRareEarthMagnetism1991}, where $\mathcal{J}(\mathbf{Q})$ refers to the Fourier transformed exchange matrix - driving home the point that \pbwo is really a quantum paramagnet and {\it not} a quantum spin liquid candidate. In the TFIM pseudospin language, we are situated in the limit of large transverse fields, meaning that the ground state can be understood as an induced ferromagnet, fully polarized by the transverse field $h = \Delta$ produced by the crystal field splitting.

Given these circumstances, linear spin-wave theory constitutes an excellent approximation and can be used to extract the "real" (non-renormalized) exchange parameters of \pbwo with high fidelity \cite{coldeaDirectMeasurementSpin2002}. But while LSWT fits to our neutron spectra establish that significant correlations in the purported kagome planes are present, they are clearly dwarfed by the dominant out-of-plane interactions, forming frustrated spin-tubes through two competing FM-AFM bonds. 
Although a sign change for (counter-)clockwise nearest-neighbor bonds might seem surprising at first glance given the visually similar Pr-O-Pr superexchange pathways, it is easy to rationalize as stemming from a difference in bond angles, which are rather close to the FM-AFM crossover expected from Kanamori-Goodenough rules \cite{goodenoughMagnetismChemicalBond}.
Furthermore, it is interesting to note here the relative differences between planar/non-planar exchange constants in the full $J = 4$ CEF model and the effective $S = 1/2$ TFIM Hamiltonian: While the former would indicate a 3D exchange topology with $\tilde{\mathcal{J}}_\Delta \sim 0.8 \tilde{\mathcal{J}}_1$, the latter points to a nearly one-dimensional physics with $\mathcal{J}_\Delta \sim 0.15 \mathcal{J}_1$. This apparent discrepancy is fully accounted for by the pseudospin projection. The easy axes between the Ising-like moments forming our breathing kagome triangles are nearly orthogonal, leading to an effective suppression proportional to $\mathbf{\Hat{z}}_i \cdot \mathbf{\Hat{z}}_j \sim 1/5$ of the in-plane couplings. While an accurate characterization of the breathing anisotropy is complicated by this suppression, we estimate $\mathcal{J}_\Delta /\mathcal{J}_\nabla \gtrsim 3$ based on our LSWT calculations - not surprising given the significant ($\sim 0.7 \text{ \AA}$) difference in bond length and the highly localized character of $4f$ orbitals.

But despite the obvious contrast between the supposed, highly degenerate breathing kagome model and the simpler, nearly 1D two-singlet physics at hand, \pbwo exhibits a number of puzzling properties that can only be understood by considering two important extensions to the pseudospin-projected TFIM Hamiltonian.

On one hand, because we are dealing with a two-singlet system, the nuclear moments are enormously enhanced by a factor $K \simeq 70$ through hyperfine coupling to the large Van-Vlack susceptibility, leading to a set of interesting $\mu$SR spectra with a peculiar field distribution, as well as an uncharacteristically large nuclear Zeeman splitting, which shows up in heat capacity measurements in applied fields.

Furthermore, due to the non-Kramers nature of the Pr$^{3+}$ ion, the material is especially susceptible to effects of structural disorder, which results in a distribution of CEF splittings $\mathcal{P}(\Delta)$. Even though our single crystal samples display only a meager $\lesssim 2 \%$ impurity concentration (in the form of Pr$^{4+}$ charge disorder and a concomitant excess of oxygen), there are dramatic effects on both the thermodynamic and dynamic properties. The heat capacity becomes $gapless$ with a power-law decay towards low temperatures, while the bulk of the dispersive crystal field excitons are restructured into a broad $continuum$ of excitations. 
We note that while both of these features are commonly provided as evidence for spin-liquid like physics, the extreme sensitivity of the magnetism to even small amounts of disorder seen here reinforces our belief that effects of structural imperfections should be carefully considered before claiming such exotic physics, especially when dealing with non-Kramers systems.
Also, the coexistence of both sharp spin-waves and a broad continuum is particularly unusual in disordered magnets - most of the known material examples with considerable structural disorder exhibit a $uniformly$ broadened spectrum \cite{hanCorrelatedImpuritiesIntrinsic2016,liCrystallineElectricFieldRandomness2017,liSpinDynamicsGriffiths2021}, not two separate components. While it is possible to interpret the sharp component in terms of local regions sufficiently removed/screened from impurities, a significant variation of impurity concentrations between the coaligned single crystals seems to be a much more likely explanation, especially considering the $\sim 1\%$ variation already found in susceptibility measurements.

There is one more subtle feature which might originate from structural disorder, namely the rapid temperature-dependent softening of the sound velocity at low fields $\mu_0 H \lesssim 1$ T. Any magnetoelastic effects associated with the Pr$^{4+}$ impurity spins should be suppressed at modest fields $g \mu_B \mu_0 H \sim k_B T$, potentially explaining the otherwise anomalously small energy scale associated with this feature. 

Finally, we compare \pbwo to other members in the rare-earth borotungstate family. The physics of Ising-like moments coupled into 1D frustrated spin-tubes arises primarily from the local environment, so given the strong structural similarities, we would expect this framework to remain intact for most of the light rare earths, as evidenced e.g. in  $R = \mathrm{Nd}$ \cite{flavianMagneticPhaseDiagram2023a}. Obvious exceptions to this would be  $R = \mathrm{Sm, Er, Tm, Yb}$ (a sign change of the Stevens factor $\alpha$ should provide a strong {\it easy plane} anisotropy \cite{jensenRareEarthMagnetism1991}), as well as Ce (the smaller $J = 5/2$ allows for significant $XY$-type exchange components despite the Ising anisotropy \cite{iwaharaExchangeInteractionMultiplets2015}). While the effects of both planar terms and unsuppressed in-plane exchanges might be interesting to study, the two-ion Hamiltonian should be decidedly more complicated due to the low symmetry. Also, judging from the twofold enhanced ordering temperature of Sm (planar) \cite{zengIncommensurateMagneticOrder2022} compared to Nd (axial) \cite{flavianMagneticPhaseDiagram2023a}, it would seem that the frustration might be partially released by the inverted anisotropy. 

In the context of the TFIM physics seen here, an exploration of the $R = \mathrm{Tb, Ho}$ members might be of particular interest. While \pbwo unfortunately ends up in a pseudospin-polarized region of phase space with a transverse field exceeding the exchanges, the crystal field scale for the latter ions should be significantly smaller \cite{jensenRareEarthMagnetism1991}. This opens up the question how the interplay between geometric/exchange frustration and transverse field-induced quantum fluctuations might modify the ground state for smaller fields $A_{cr} \gtrsim 1$ - especially whether a strongly correlated quantum disordered phase can remain stable in an extended regime $\mathcal{J}_{\rm MF} \gtrsim h$. Potential structural disorder should also be more subdued in these materials, though its effects might even be beneficial, having been suggested to stabilize a quantum spin liquid phase in certain frustrated systems \cite{savaryDisorderInducedQuantumSpin2017}.

\section{Conclusion}

Despite the promise of spin-liquid physics on a breathing kagome lattice, \pbwo turns out to be a relatively simple quantum paramagnet. While it does feature significant frustration inherent from both the lattice topology and an FM-AFM exchange competition, the ground state properties are dictated by a crystal field splitting $\Delta$ of the lowest quasi-doublet. This results in a simple effective Hamiltonian of a frustrated Ising model in a transverse magnetic field, which can be modeled numerically to astounding success. Our neutron spectroscopy results demonstrate that the dominant interactions are those {\it perpendicular to} the kagome planes, confirming that the $R_3$BWO$_9$ family of quantum antiferromagnets is best described as an arrangement of 1D frustrated triangular spin-tubes, moderately coupled through the planar breathing kagome exchanges. The two-singlet nature of \pbwo makes it especially susceptible to the influence of hyperfine enhanced nuclear moments and weak structural disorder, both of which have dramatic effects on the static and dynamic properties, including an unusual coexistence of both sharp spin-waves and broad continuum excitations in an otherwise simple model system. Frustrated quantum Ising models remain a promising avenue in the ongoing hunt for novel emergent quantum magnetic phases, although the desired parametric regime of frustrated exchanges and strong quantum fluctuations continues to prove rather difficult to engineer.

\section{Acknowledgements}

We thank Prof. Titus Neupert and Nikita Astrakhantsev (University of Zürich) for illuminating discussions. This work is supported by a MINT grant of the Swiss National Science Foundation. We acknowledge the support of the HLD at HZDR, member of the European Magnetic Field Laboratory (EMFL), and the W\"{u}rzburg-Dresden Cluster of Excellence on Complexity and Topology in Quantum Matter - ct.qmat (EXC 2147, project ID 390858490). We thank Sergei Zherlitsyn and Yurii Skourski (HZDR) for help with the pulsed-field experiments. The neutron experiment at AMATERAS in J-PARC/MLF was performed under a user program (Proposal No. 2022A0022). We acknowledge the beam time allocation at PSI (EIGER id: 20211049, HRPT id: 20230203).

\appendix
\section{Effective Point-Charge Model}
\label{sec:A1}

Following the method pioneered by Hutchings \cite{hutchingsPointChargeCalculationsEnergy1964}, we can approximate the crystal field Hamiltonian Eq.~(\ref{eq:CEF}) by treating the ligands surrounding our magnetic ions as point charges $q_i$ and simply calculating the electrostatic potential. The CEF parameters $B^m_n$ can be expressed as
\begin{gather}
	B^m_n = -\sum_i q_i \gamma^{nm}_i C^m_n \langle r^n \rangle \theta_n,
\end{gather}
where both $\gamma^{nm}_i$ and $C^m_n$ are tabulated factors of the tesseral harmonics, $\langle r^n \rangle$ is the expectation value of the radial wavefunction \cite{EdvardsonElectrostaticCEF1998} and $\theta_n$ represent the Stevens factors. As such, we can express Eq.~(\ref{eq:CEF}) purely in terms of the charges $q_i$ corresponding to the three inequivalent O$^{2-}$ ligands, dramatically reducing the number of free parameters.

\begin{table}[tbp]
	\caption{CEF parameters of the effective point-charge model, in the global $(a,b^{*},c)$ frame and the local frame of each ion.}
	\begin{tabularx}{0.45\textwidth}{YYY} \\
		\toprule\toprule
		$B_n^m$ ($10^{-2}$ meV) & Global Frame & Local Frame \\\midrule
		$B_2^{-2}$ & -45.732 & -24.513 \\
		$B_2^{-1}$ & -464.458 & -439.061 \\
		$B_2^0$ & 30.383 & 45.214 \\
		$B_2^1$ & -358.590 & -288.734 \\
		$B_2^2$ & -86.931 & -262.987 \\
		$B_4^{-4}$ & -0.776 & -3.805 \\
		$B_4^{-3}$ & -11.341 & 6.819 \\
		$B_4^{-2}$ & -0.786 & 0.777 \\
		$B_4^{-1}$ & 6.168 & 11.124 \\
		$B_4^0$ & 1.121 & 1.708 \\
		$B_4^1$ & -7.957 & -5.627 \\
		$B_4^2$ & -1.515 & -2.545 \\
		$B_4^3$ & -1.005 & -15.699 \\
		$B_4^4$ & 2.920 & -3.648 \\
		$B_6^{-6}$ & -0.114 & 0.119 \\
		$B_6^{-5}$ & 0.099 & 0.264 \\
		$B_6^{-4}$ & 0.068 & 0.120 \\
		$B_6^{-3}$ & 0.020 & -0.044 \\
		$B_6^{-2}$ & -0.057 & -0.066 \\
		$B_6^{-1}$ & 0.066 & 0.096 \\
		$B_6^0$ & 0.002 & 0.006 \\
		$B_6^1$ & -0.060 & -0.007 \\
		$B_6^2$ & 0.023 & -0.044 \\
		$B_6^3$ & 0.008 & 0.051 \\
		$B_6^4$ & -0.132 & 0.172 \\
		$B_6^5$ & -0.444 & 0.451 \\
		$B_6^6$ & -0.024 & 0.025 \\
		\toprule
	\end{tabularx}
	\label{tab:CEF_par}
\end{table}

\begin{table}[tbp]
	\caption{Single-ion wavefunctions of the Pr$^{3+}$ ground state quasi-doublet obtained from point-charge calculations (top) in the global crystallographic frame $(a,b^{*},c)$ and (bottom) in the local frame of each ion, related to the global one given above through Euler angle rotations $(\alpha,\beta,\gamma) \approx (10.0^{\circ},46.5^{\circ},-1.1^{\circ})$ in $z$-$y$-$z$ convention.}
	\begin{tabularx}{0.45\textwidth}{c|YY}
		\multicolumn{3}{c}{} \\
		\toprule\toprule
		& $\ket{\psi_{0}}$ & $\ket{\psi_{1}}$ \\
		\midrule
		$\ket{-4}$ & $0.252$ & $0.269$ \\
		$\ket{-3}$ & $-0.468-0.103i$ & $-0.458-0.131i$ \\
		$\ket{-2}$ & \;\;\;$0.170+0.396i$ & \;\;\;$0.148+0.356i$ \\
		$\ket{-1}$ & \;\;\;$0.038-0.099i$ & \;\;\;$0.087-0.203i$ \\
		$\ket{0}$ & $-0.110+0.096i$ & \;\;\;$0.049+0.058i$ \\
		$\ket{+1}$ & $-0.103+0.025i$ &  \;\;\;$0.216-0.050i$ \\
		$\ket{+2}$ & $-0.371-0.219i$ & \;\;\;$0.325+0.208i$ \\
		$\ket{+3}$ & $-0.042-0.477i$ & \;\;\;$0.049+0.474i$ \\
		$\ket{+4}$ & \;\;\;$0.032-0.250i$ & $-0.047+0.265i$ \\
		\toprule
	\end{tabularx}
	\begin{tabularx}{0.45\textwidth}{c|YY}
		\multicolumn{3}{c}{} \\
		\toprule\toprule
		& $\ket{\psi_{0}}$ & $\ket{\psi_{1}}$ \\
		\midrule
		$\ket{-4}$ & \;\;\;$0.678+0.001i$ & $-0.686+0.001i$ \\
		$\ket{-3}$ & $-0.051-0.020i$ & \;\;\;$0.044+0.045i$ \\
		$\ket{-2}$ & $-0.101-0.041i$ & \;\;\;$0.108+0.010i$ \\
		$\ket{-1}$ & $-0.114-0.108i$ & \;\;\;$0.035+0.073i$ \\
		$\ket{0}$ & $0.029$ & $0.123i$ \\
		$\ket{+1}$ & \;\;\;$0.114-0.107i$ & \;\;\;$0.035-0.073i$ \\
		$\ket{+2}$ & $-0.101+0.041i$ & $-0.108+0.011i$ \\
		$\ket{+3}$ & \;\;\;$0.051-0.020i$ & \;\;\;$0.044-0.045i$ \\
		$\ket{+4}$ & $0.678$ & $0.686$ \\
		\toprule
	\end{tabularx}
	\label{tab:wavefunc}
\end{table}

By diagonalizing this CEF Hamiltonian, we obtain the energies $E_i$ and wavefunctions $\ket{\psi_i}$ of a single Pr$^{3+}$ ion, which may be used to fit and compare to experimental data, namely INS and magnetometry. The powder averaged neutron scattering intensity in the dipole approximation is given as
\begin{eqnarray}
	\mathcal{I} (Q,\omega) = A F^2(Q) e^{-2W(Q)} \sum_{ij\alpha} p_i |\bra{\psi_i} \hat{J}_\alpha \ket{\psi_j}|^2 \nonumber \\*
	\times \delta(\hbar\omega + E_i - E_j) \qquad\qquad\qquad
\end{eqnarray}
where $A$ is a scale factor, $F(Q)$ represents the magnetic form factor, $e^{-2W(Q)}$ is the Debye-Waller factor and $p_i$ corresponds to the Boltzmann weights. In practice, the delta function is replaced by a Voigt profile to account for the instrumental resolution and the finite lifetime of excitations, while an intensity correction is introduced to deal with the energy transfer-dependent resolution volume of the spectrometer \cite{PopoviciResolution1975}. On the other hand, the single-ion magnetization is calculated as
\begin{gather}
	\label{eq:Magnetiz}
	M^{\rm CEF}_\alpha (\mathbf{H},T) = g_J \mu_{\rm B} \sum_i p_i \bra{\psi_i} \hat{J}_\alpha \ket{\psi_i} 
\end{gather}
where the uniform susceptibility can be simply evaluated numerically as
\begin{gather}
	\chi_{\alpha \beta} = \frac{\partial M_\alpha}{\partial H_\beta}.
\end{gather}
To recover the bulk magnetic response observed in experiment, these quantities are averaged over all six ions in the unit cell (related by sixfold rotations $\mathbf{R}_i$ in the hexagonal plane), i.e. 
\begin{gather}
	\label{eq:Mavg}
	\bar{M}^{\rm CEF}_\alpha = \frac{1}{6} \sum_i M^{\rm CEF}_\alpha (\mathbf{R}_i \cdot \mathbf{H},T).
\end{gather}
In the low temperature regime where two-ion correlations start to become relevant, corrections to Eq.~(\ref{eq:Magnetiz}-\ref{eq:Mavg}) can be mostly accounted for by a Weiss molecular field. Each ion experiences a local field $\mathbf{H}_{\rm Weiss}$ proportional to the magnetization of its neighbors, determined self-consistently using
\begin{align}
	\qquad\qquad \bar{\mathbf{M}}^{\rm MF} &= \bar{\mathbf{M}}^{\rm CEF} (\mathbf{H} + \mathbf{H}_{\rm Weiss},T) \quad {\rm where} \nonumber \\*
	\mathbf{H}_{\rm Weiss} &= -\frac{2 \mathcal{J}_{\rm MF} V_0}{({\bf g}_{\rm avg} \mu_{\rm B})^2}\bar{\mathbf{M}}^{\rm MF}.
\end{align}
Here, $V_0$ is the unit cell volume and $\mathbf{g}_{\rm avg}$ corresponds to the unit cell-averaged $g$-tensor, leaving only the mean-field exchange constant $\mathcal{J}_{\rm MF} = \frac{1}{2} \sum_{ij} \mathcal{J}_{ij}$ as a tuning parameter to deal with interactions.

An effective point-charge fit to the combined INS and magnetometry data was carried out using a least-squares refinement with four fit parameters: The charges $q_i$ of the three inequivalent O$^{2-}$ ligands caging the Pr$^{3+}$ ions, as well as the mean field exchange constant $\mathcal{J}_{\rm MF} \simeq 0.33$ meV. The CEF parameters $B_n^m$ obtained from the fitting procedure are presented in Table \ref{tab:CEF_par}, while the wavefunctions of the lowest quasi-doublet are given in Table \ref{tab:wavefunc}. The optimized CEF spectrum consists of nine isolated singlets, with excitations at $E_{i}^{\rm CEF} = 1.7$, 37.0, 42.6, 58.0, 74.8, 91.8, 101.1 and 105.1 meV. Due to the $\sim 400$ K gap to higher CEF excitations, the lowest two levels form a well isolated quasi-doublet ground state with a small splitting $\Delta \sim 1.7$ meV. Nearly all of the neutron scattering intensity is concentrated on this transition, with further modes suppressed by a factor $\gtrsim 10^{-3}$ and hence not visible in experiment above noise level. This is exacerbated at high temperatures, where the combination of lifetime broadening effects and strongly increased phonon population makes it increasingly difficult to identify CEF transitions. 

\begin{table*}[tbp]
	\caption{Crystal structure parameters of \pbwo at $T = 1.5$ K determined through neutron diffraction on powders. The $z$-coordinate of $^{11}$B remains fixed at 0 during the refinement, as the $P6_3$ structure has no preferred origin along that axis.}
	
	\begin{tabularx}{0.8\textwidth}{YYYYYY} \\
		\toprule\toprule
		Atom & $x$ & $y$ & $z$ & Occ. & Equiv. $B_\mathrm{iso}$ \\
		\midrule
		Pr & 0.0854(3) & 0.7245(3) & 0.3460(7) & 1.00(1) & 0.27(5) \\
		$^{11}$B & 0 & 0 & 0 & 0.32(1) & 0.49(9) \\
		W & 0.3333 & 0.6667	& 0.8819(9) & 0.33(1) & 0.23(7) \\
		O(1) & 0.0459(2) & 0.8689(2) & 0.9940(6) & 1.06(1) & 0.65(6) \\
		O(2) & 0.1413(2) & 0.5192(2) & 0.1067(6) & 1.07(1) & 0.49(4) \\
		O(3) & 0.1931(2) & 0.4759(2) & 0.6724(6) & 1.06(1) & 0.98(8) \\
		\toprule
	\end{tabularx}
	\label{tab:Structure}
\end{table*}

\section{Pseudospin Projection}
\label{sec:A2}

Due to the large $\sim 400$ K gap to higher CEF excitations, only the quasi-doublet states defined in Eq.~(\ref{eq:Doublet}) are populated at low temperatures and the thermodynamic/magnetic properties are fully determined by these low-energy degrees of freedom. Therefore, it becomes useful to "integrate out" the irrelevant states and calculate the projection of the magnetic moments $\hat{\mathbf{J}}$ onto the $2 \times 2$ quasi-doublet subspace spanned by $\ket{\pm} \simeq \ket{J_z = \pm 4}$, given as
\begin{gather}
	\hat{\mathcal{P}} \hat{J}^z \hat{\mathcal{P}} = \frac{g_{zz}}{g_J} \hat{S}^z, \qquad
	\hat{\mathcal{P}} \hat{J}^{\pm} \hat{\mathcal{P}} = 0.
\end{gather}
Here, $\hat{\mathcal{P}} = \ket{+}\bra{+} + \ket{-}\bra{-}$ defines the ground state projector and the local pseudospin operators $\hat{\mathbf{S}}$ are constructed in the standard framework \cite{rauFrustratedQuantumRareEarth2019}
\begin{gather}
	\hat{S}^z \equiv \frac{\ket{+}\bra{+} - \ket{-}\bra{-}}{2}, \qquad
	\hat{S}^\pm \equiv \ket{\pm}\bra{\mp}
\end{gather}
from the two thermally accessible crystal field states. We note that our choice of basis $\ket{\pm}$ conforms to the convention common to pyrochlore oxides \cite{rauFrustratedQuantumRareEarth2019, martinDisorderQuantumSpin2017a}. In principle, an equivalent formulation in terms of the $\ket{0}$ and $\ket{1}$ states is possible and would correspond simply to swapping our notation of $\hat{S}^z$ and $\hat{S}^x$ throughout the text. With this, the total magnetic dipole moment operator can be written as
\begin{gather}
	\hat{\bm{\mu}} = -g_J \mu_{\rm B} \hat{\mathcal{P}} \hat{\mathbf{J}} \hat{\mathcal{P}} = -g_{zz} \mu_{\rm B} \hat{\mathbf{z}} \hat{S}^z
\end{gather}
where $g_{zz} = 2 g_J \bra{1} \hat{J}^z \ket{0} \approx 6.09$ is the longitudinal $g$-factor and $g_\perp = 0$ (as is the case for all non-Kramers (quasi-)doublets). Despite the perfect axial single-ion anisotropy, upon averaging over all six ions in the unit cell we obtain $g_c \approx 2.89$ and $g_{ab} \approx 1.60$, in excellent agreement with experiment. Only $\hat{S}^z$ contributes to the magnetic dipole moment, represented as either "up" or "down". On the other hand, the transverse pseudospin components show up in the projected crystal field Hamiltonian
\begin{gather}
	\hat{\mathcal{P}} \mathcal{H}_{\rm CEF} \hat{\mathcal{P}} = \Delta \hat{S}^x,
\end{gather}
acting as a transverse field $h = \Delta$ on the otherwise Ising-like moments and mixing our $\ket{\pm}$ basis into the quasi-doublet ground states given in Eq.~(\ref{eq:Doublet}). Since those two states are singlets, we have $\bra{0} \hat{\mathbf{J}} \ket{0} = \bra{1} \hat{\mathbf{J}} \ket{1} = 0$ and only the transverse matrix elements $\bra{1} \hat{J}_z \ket{0} \approx 3.81 \sim J$ are nonzero - meaning that the neutron scattering cross section is purely {\it inelastic}. All in all, we can establish a direct mapping between the low-energy physics of our full $J = 4$ CEF Hamiltonian Eq.~(\ref{eq:H_full}) and that of Ising-like pseudospin $S = 1/2$ moments in a strong transverse field (TFIM) as in Eq.~(\ref{eq:Hamilton_2lvl}).

\section{Crystal Structure}

Rietveld refinements based on the low-temperature neutron powder diffraction experiment discussed in Sec.~\ref{sec:Disorder} were carried out in the hexagonal $P6_3$ space group. The refined lattice parameters are $a = 8.7210(1) \text{ \AA}$ and $c = 5.5049(1) \text{ \AA}$, while the atomic coordinates, site occupancy and equivalent isotropic displacement parameters \cite{truebloodAtomicDisplacementParameter1996} are provided in Table \ref{tab:Structure}. The coordinates obtained in this way are deemed to be more suitable for point-charge calculations than other refinements available in the literature \cite{ashtarNewFamilyDisorderFree2020,krutkoStructuresNonlinearHexagonal2006}, as the positions of the light O$^{2-}$ ligands directly responsible for the CEF environment are extremely difficult to refine accurately through x-ray diffraction, especially in the presence of the heavy tungsten. All structural considerations and bond angles/distances discussed in Sec.~\ref{ssec:mat} are based on these parameters.

\bibliography{R3BWO9_bib}

\end{document}